\documentclass[a4paper]{article}
\usepackage{amsmath,amssymb,graphicx,color}
\definecolor{gray}{rgb}{0.8,0.8,0.8}
\definecolor{brown}{rgb}{1,0.5,0.5}
\definecolor{purple}{rgb}{0.5,0.5,1}
\newlength{\myboxw}
\newlength{\mygridw}
\newcounter{myc}
\newcommand{\1}[2]{\linethickness{\myboxw}\color{black}\put(#1,#2){\put(0,0.5){\line(1,0){1}}}}
\newcommand{\0}[2]{\linethickness{\myboxw}\color{white}\put(#1,#2){\put(0,0.5){\line(1,0){1}}}}
\newcommand{\A}[2]{\linethickness{\myboxw}\color{brown}\put(#1,#2){\put(0,0.5){\line(1,0){1}}}}
\newcommand{\B}[2]{\linethickness{\myboxw}\color{purple}\put(#1,#2){\put(0,0.5){\line(1,0){1}}}}

\newcommand{\mygrid}[2]{\linethickness{\mygridw}\color{gray}\put(0,0){%
\setcounter{myc}{0}
\loop\ifnum\value{myc}<#2
  \put(0,\value{myc}){\line(1,0){#1}}
  \stepcounter{myc}
\repeat
\put(0,\value{myc}){\line(1,0){#1}}
\setcounter{myc}{0}
\loop\ifnum\value{myc}<#1
  \put(\value{myc},0){\line(0,1){#2}}
  \stepcounter{myc}
\repeat
\put(\value{myc},0){\line(0,1){#2}}}
}
\everymath{\displaystyle}
\begin{document}
\title{Max-Plus Generalization of Conway's Game of Life}
\author{Kotaro Sakata, Yuta Tanaka and Daisuke Takahashi\\
Department of Pure and Applied Mathematics, Waseda University,\\
3-4-1, Okubo, Shinjuku-ku, Tokyo, 169-8555, Japan}
\date{}
\maketitle
%%%%%%%%%%%%%%%%%%%%%%%%%%%%%%%%%%%%%%%%%%%%%%%%%%%%%%%%%%%%
\begin{abstract}
  We propose a max-plus equation which includes Conway's Game of Life (GoL) as a special case.  There are some special solutions to the equation which include and unify those to GoL.  Moreover, the multi-value extension of GoL is derived from the equation and the behavior of solutions is discussed.
\end{abstract}
\section{Introduction}
%%%%%%%%%%%%%%%%%%%%%%%%%%%%%%%%%%%%%%%%%%%%%%%%%%%%%%%%%%%%
  Conway's Game of Life (GoL) is a binary cellular automaton (CA) and expresses a kind of population ecology\cite{gardner,wolfram}.  It is an evolution game using a two-dimensional orthogonal grid of cells and each cell has either of two states, alive or dead.  The evolution rule for the discrete generation is defined as follows.
\begin{enumerate}
\item
  Birth: If there are just 3 live cells in the Moore neighborhood of a dead cell, the dead cell changes to the live cell at the next generation.
\item
  Survival: If there are 2 or 3 live cells in the Moore neighborhood of a live cell, it is alive at the next generation.
\item
  Death: Otherwise, the cell is dead at the next generation.
\end{enumerate}
Let us assume values of two states, 1 for alive and 0 for dead.  If $u_{ij}^n$ denotes the value at $(i,j)$ cell of the generation $n$, the above evolution rule can be transcribed into the evolution equation,
\begin{equation}  \label{GoL}
  u_{ij}^{n+1}=
\begin{cases}
  1 & \text{($(u_{ij}^n,s_{ij}^n)=(0,3)$, $(1,2)$ or $(1,3)$)} \\
  0 & \text{(otherwise)}
\end{cases},
\end{equation}
where $s_{ij}^n=u_{i-1j-1}^n+u_{ij-1}^n+u_{i+1j-1}^n+u_{i-1j}^n+u_{i+1j}^n+u_{i-1j+1}^n+u_{ij+1}^n+u_{i+1j+1}^n$.  Specific solutions to GoL have been vastly searched and listed.  There are various types of evolution of solutions and they are analogous to activities of life.\par
  In this paper, we propose an extended model of GoL using max-plus operation.  Max-plus operation is based on max-plus algebra which is a commutative semiring defined by addition `max' and multiplication `+'\cite{gaubert}.  It is used for the description of discrete event systems and utilized to analyze the dynamics with max-plus linear system theory based on the Perron-Frobenius theory\cite{baccelli}.  It is also obtained by ultradiscretizing the difference equations through the limiting procedure,
\begin{equation}  \label{ultra}
  \lim_{\varepsilon\to+0}\varepsilon\log(e^{A/\varepsilon}+e^{B/\varepsilon})=\max(A,B).
\end{equation}
Tokihiro et. al. found that the binary CA called `box and ball system' is obtained by ultradiscretizing the discrete soliton equation through the above limit and showed that the solutions to the box and ball system giving soliton interactions among groups of balls can be derived by the same limit of multi-soliton solutions to the discrete equation\cite{tokihiro,takahashi}.\par
  In the above context, we can consider that max-plus expression proposes novel viewpoint and mathematical tools to pure digital systems like CA.  We propose a max-plus equation with a continuous dependent variable in which GoL is embedded as a special case.  There exist real-valued exact solutions to the equation and they include and unify those to GoL.  Contents of this paper are as follows.  In Section~\ref{sec:definition}, the max-plus equation extended from GoL is proposed.  In Section~\ref{sec:solutions}, special solutions to the max-plus equation and their relations to those to GoL are shown.  In Section~\ref{sec:multi-value}, we show the multi-value CA obtained from the max-plus equation and discuss the behavior of solutions.  In Section~\ref{sec:conclusion}, we give concluding remarks.
%%%%%%%%%%%%%%%%%%%%%%%%%%%%%%%%%%%%%%%%%%%%%%%%%%%%%%%%%%%%
\section{Definition of MaxLife}  \label{sec:definition}
\begin{figure}[hbtp]
\begin{center}
  \includegraphics[scale=0.8]{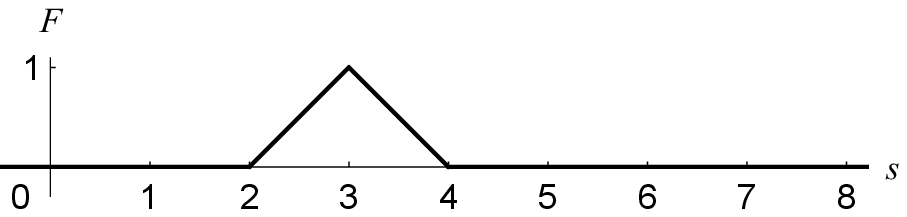} \\
  (a) $F(0,s)$ \\
  \includegraphics[scale=0.8]{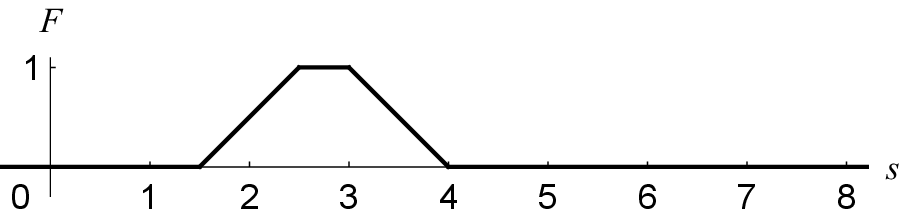} \\
  (b) $F(0.5,s)$\\
  \includegraphics[scale=0.8]{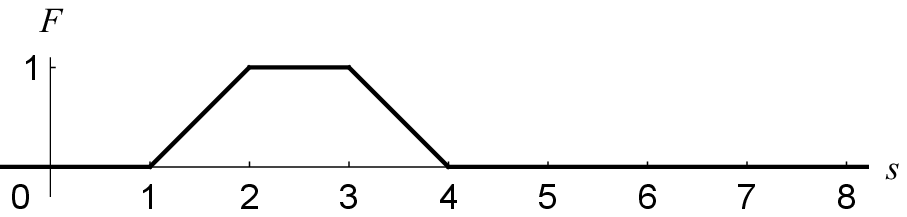} \\
  (c) $F(1,s)$
\end{center}
  \caption{Graphs of $F(u,s)$.}
  \label{fig:F}
\end{figure}

%%%%%%%%%%%%%%%%%%%%%%%%%%%%%%%%%%%%%%%%%%%%%%%%%%%%%%%%%%%%
  Let us consider the following evolution equation using operators $\max$, $+$ and $-$.
\begin{equation}  \label{maxlife}
\begin{aligned}
  & u_{ij}^{n+1}=F(u_{ij}^n,s_{ij}^n),\\
\end{aligned}
\end{equation}
where $i$ and $j$ are integer space coordinates, $n$ integer time, $s_{ij}^n$ the sum of eight $u$'s in the Moore neighborhood,
\begin{equation*}
\begin{aligned}
  s_{ij}^n
  &=u_{i-1j-1}^n+u_{ij-1}^n+u_{i+1j-1}^n+u_{i-1j}^n+u_{i+1j}^n \\
  &\qquad+u_{i-1j+1}^n+u_{ij+1}^n+u_{i+1j+1}^n,
\end{aligned}
\end{equation*}
and $F(u,s)$ is defined by
\begin{equation*}
\begin{aligned}
  F(u,s)&=\max(0,u+s-2)-\max(0,u+s-3) \\
        &\qquad-\max(0,s-3)+\max(0,s-4).
\end{aligned}
\end{equation*}
If $0\le u\le1$, we can easily show $0\le F(u,s)\le1$.  Figure~\ref{fig:F} shows the graphs of $F(0,s)$, $F(0.5,s)$ and $F(1,s)$.
\par
  Consider the initial value problem for equation (\ref{maxlife}) and assume $n=0$ is an initial time.  If we set the initial data $u_{ij}^0$ to satisfy $0\le u_{ij}^0\le 1$ for any $i$ and $j$, then any $u_{ij}^n$ ($n>0$) also since $0\le F(u,s)\le1$.  Moreover, if we assume $u_{ij}^n$ at a certain $n$ for any $i$ and $j$ takes either of the values 0 and 1, $s_{ij}^n$ is one of the nine integer values from 0 to 8.  Then the value of RHS of equation (\ref{maxlife}) is also 0 or 1 considering the graphs of $F(0,s)$ and $F(1,s)$.  Therefore, the value of solution $u_{ij}^n$ to equation (\ref{maxlife}) can be closed in the binary set $\{0,1\}$ if the initial data $u_{ij}^0$ is.  Then the evolution equation (\ref{maxlife}) becomes equivalent to equation (\ref{GoL}) considering the profiles of $F(0,s)$ and $F(1,s)$.  Thus equation (\ref{maxlife}) includes the rule of GoL as a special case.  We call the evolution system defined by equation (\ref{maxlife}) `MaxLife' in this meaning and discuss the behavior of its real-valued solutions closed in the range of $[0,1]$ relating them to binary solutions closed in the range of $\{0,1\}$ which are also solutions to GoL.
%%%%%%%%%%%%%%%%%%%%%%%%%%%%%%%%%%%%%%%%%%%%%%%%%%
\section{Special solutions to MaxLife}  \label{sec:solutions}
%%%%%%%%%%%%%%%%%%%%%%%%%%%%%%%%%%%%%%%%%%%%%%%%%%
  In this section, we show the special solutions to MaxLife.  Since it is difficult to solve equation (\ref{maxlife}) in a systematic way,  we assume dimensions, symmetry and period of solutions within the background $u=0$.  Solutions shown below is confined in $4\times4$ region at most, and static, periodic or moving stably.  They reduce to binary solutions to GoL in a special case and almost solutions unify two or more solutions to GoL.
%%%%%%%%%%%%%%%%%%%%%%%%%%%%%%%%%%%%%%%%%%%%%%%%%%
\subsection{Static solution confined in $2\times2$ region (`Block')}
%%%%%%%%%%%%%%%%%%%%%%%%%%%%%%%%%%%%%%%%%%%%%%%%%%
  The first example is a static solution confined in $2\times2$ region,
\begin{equation*}
\begin{array}{l}
  \ldots0\,0\,0\,0\,0\,0\ldots\\
  \ldots0\,0\,c\,d\,0\,0\ldots\\
  \ldots0\,0\,a\,b\,0\,0\ldots\\
  \ldots0\,0\,0\,0\,0\,0\ldots\\
\end{array},
\end{equation*}
where $a$, $b$, $c$ and $d$ are all real constants from 0 to 1.  The region outside the shown is $u=0$.  Substituting the above solution to equation (\ref{maxlife}), we obtain one of equations,
\begin{equation*}
\begin{aligned}
  a&=\max(0,a+b+c+d-2)-\max(0,a+b+c+d-3)\\
  &\qquad-\max(0,b+c+d-3)+\max(0,b+c+d-4).
\end{aligned}
\end{equation*}
Since $a$, $b$, $c$, $d\in[0,1]$, the above equation reduces to
\begin{equation*}
  a=\max(0,a+b+c+d-2)-\max(0,a+b+c+d-3)
\end{equation*}
The RHS is symmetric about constants and the following condition is obtained considering other equations.
\begin{equation*}
  a=b=c=d,\qquad a=\max(0,4a-2)-\max(0,4a-3).
\end{equation*}
Solving this condition, we have
\begin{equation*}
  a=b=c=d=0\text{\ or\ }\frac{2}{3}\text{\ or\ }1.
\end{equation*}
The above values give a trivial solution ($a=0$), a non-integer solution ($a=2/3$) and a binary solution to GoL called `block' ($a=1$).
%%%%%%%%%%%%%%%%%%%%%%%%%%%%%%%%%%%%%%%%%%%%%%%%%%
\subsection{`Blinker' type of solution}
\begin{figure}[hbtp]
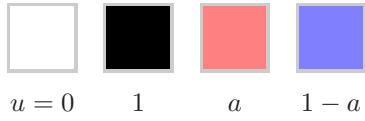

\setlength{\fboxrule}{1pt}
\setlength{\fboxsep}{12pt}
\begin{center}
$\begin{array}{cccc}
  \fcolorbox{gray}{white}{}
&
  \fcolorbox{gray}{black}{}
&
  \fcolorbox{gray}{brown}{}
&
  \fcolorbox{gray}{purple}{}
\medskip
\\
 u=0 & 1 & a & 1-a
\end{array}$
\end{center}
\caption{Four colored cells denoting the values of $u$.}
\label{fig:color}
\end{figure}

\begin{figure}[hbtp]
\begin{center}
$\vcenter{\hbox{\begin{picture}(5,5)
\0{0}{4}\0{1}{4}\0{2}{4}\0{3}{4}\0{4}{4}
\0{0}{3}\0{1}{3}\A{2}{3}\0{3}{3}\0{4}{3}
\0{0}{2}\B{1}{2}\1{2}{2}\B{3}{2}\0{4}{2}
\0{0}{1}\0{1}{1}\A{2}{1}\0{3}{1}\0{4}{1}
\0{0}{0}\0{1}{0}\0{2}{0}\0{3}{0}\0{4}{0}
\mygrid{5}{5}
\end{picture}}}
\leftrightarrow
\vcenter{\hbox{\begin{picture}(5,5)
\0{0}{4}\0{1}{4}\0{2}{4}\0{3}{4}\0{4}{4}
\0{0}{3}\0{1}{3}\B{2}{3}\0{3}{3}\0{4}{3}
\0{0}{2}\A{1}{2}\1{2}{2}\A{3}{2}\0{4}{2}
\0{0}{1}\0{1}{1}\B{2}{1}\0{3}{1}\0{4}{1}
\0{0}{0}\0{1}{0}\0{2}{0}\0{3}{0}\0{4}{0}
\mygrid{5}{5}
\end{picture}}}$
\end{center}
\caption{`Blinker' type of solution.}
\label{fig:max blinker}
\end{figure}
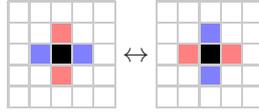

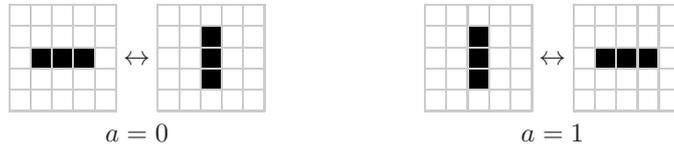
\begin{figure}[hbtp]
\begin{center}
$\begin{array}{ccc}
\vcenter{\hbox{\begin{picture}(5,5)
\0{0}{4}\0{1}{4}\0{2}{4}\0{3}{4}\0{4}{4}
\0{0}{3}\0{1}{3}\0{2}{3}\0{3}{3}\0{4}{3}
\0{0}{2}\1{1}{2}\1{2}{2}\1{3}{2}\0{4}{2}
\0{0}{1}\0{1}{1}\0{2}{1}\0{3}{1}\0{4}{1}
\0{0}{0}\0{1}{0}\0{2}{0}\0{3}{0}\0{4}{0}
\mygrid{5}{5}
\end{picture}}}
\leftrightarrow
\vcenter{\hbox{\begin{picture}(5,5)
\0{0}{4}\0{1}{4}\0{2}{4}\0{3}{4}\0{4}{4}
\0{0}{3}\0{1}{3}\1{2}{3}\0{3}{3}\0{4}{3}
\0{0}{2}\0{1}{2}\1{2}{2}\0{3}{2}\0{4}{2}
\0{0}{1}\0{1}{1}\1{2}{1}\0{3}{1}\0{4}{1}
\0{0}{0}\0{1}{0}\0{2}{0}\0{3}{0}\0{4}{0}
\mygrid{5}{5}
\end{picture}}}
&
\qquad\qquad
&
\vcenter{\hbox{\begin{picture}(5,5)
\0{0}{4}\0{1}{4}\0{2}{4}\0{3}{4}\0{4}{4}
\0{0}{3}\0{1}{3}\1{2}{3}\0{3}{3}\0{4}{3}
\0{0}{2}\0{1}{2}\1{2}{2}\0{3}{2}\0{4}{2}
\0{0}{1}\0{1}{1}\1{2}{1}\0{3}{1}\0{4}{1}
\0{0}{0}\0{1}{0}\0{2}{0}\0{3}{0}\0{4}{0}
\mygrid{5}{5}
\end{picture}}}
\leftrightarrow
\vcenter{\hbox{\begin{picture}(5,5)
\0{0}{4}\0{1}{4}\0{2}{4}\0{3}{4}\0{4}{4}
\0{0}{3}\0{1}{3}\0{2}{3}\0{3}{3}\0{4}{3}
\0{0}{2}\1{1}{2}\1{2}{2}\1{3}{2}\0{4}{2}
\0{0}{1}\0{1}{1}\0{2}{1}\0{3}{1}\0{4}{1}
\0{0}{0}\0{1}{0}\0{2}{0}\0{3}{0}\0{4}{0}
\mygrid{5}{5}
\end{picture}}}
\smallskip\\
a=0 && a=1
\end{array}$
\end{center}
\caption{`Blinker' of GoL.}
\label{fig:blinker}
\end{figure}

%%%%%%%%%%%%%%%%%%%%%%%%%%%%%%%%%%%%%%%%%%%%%%%%%%
Solutions from this subsection are shown schematically as figures without proof.  The four colored cells shown in Figure~\ref{fig:color} are used to denote the values of $u$ where $a$ is any constant from 0 to 1.
The `blinker' type of solution is shown in Figure~\ref{fig:max blinker}.  This solution is periodic with period 2.  The double arrow `$\leftrightarrow$' means the state at next time of the left (right) state is the right (left).
Since the `blinker' of GoL is obtained in the case of $a=0$ and 1, the solution in Figure~\ref{fig:max blinker} includes two configurations of `blinker' rotated 90 degree to each other as shown in Figure~\ref{fig:blinker}.
%%%%%%%%%%%%%%%%%%%%%%%%%%%%%%%%%%%%%%%%%%%%%%%%%%
\subsection{`Clock' type of solution}  \label{subsec:clock}
\begin{figure}[hbtp]
\begin{center}
$\begin{array}{ll}
\vcenter{\hbox{\begin{picture}(6,6)
\0{0}{5}\0{1}{5}\0{2}{5}\0{3}{5}\0{4}{5}\0{5}{5}
\0{0}{4}\0{1}{4}\0{2}{4}\B{3}{4}\0{4}{4}\0{5}{4}
\0{0}{3}\B{1}{3}\A{2}{3}\1{3}{3}\0{4}{3}\0{5}{3}
\0{0}{2}\0{1}{2}\1{2}{2}\A{3}{2}\B{4}{2}\0{5}{2}
\0{0}{1}\0{1}{1}\B{2}{1}\0{3}{1}\0{4}{1}\0{5}{1}
\0{0}{0}\0{1}{0}\0{2}{0}\0{3}{0}\0{4}{0}\0{5}{0}
\mygrid{6}{6}
\end{picture}}}
\leftrightarrow
\vcenter{\hbox{\begin{picture}(6,6)
\0{0}{5}\0{1}{5}\0{2}{5}\0{3}{5}\0{4}{5}\0{5}{5}
\0{0}{4}\0{1}{4}\B{2}{4}\0{3}{4}\0{4}{4}\0{5}{4}
\0{0}{3}\0{1}{3}\A{2}{3}\1{3}{3}\B{4}{3}\0{5}{3}
\0{0}{2}\B{1}{2}\1{2}{2}\A{3}{2}\0{4}{2}\0{5}{2}
\0{0}{1}\0{1}{1}\0{2}{1}\B{3}{1}\0{4}{1}\0{5}{1}
\0{0}{0}\0{1}{0}\0{2}{0}\0{3}{0}\0{4}{0}\0{5}{0}
\mygrid{6}{6}
\end{picture}}}
&
\vcenter{\hbox{\begin{picture}(6,6)
\0{0}{5}\0{1}{5}\0{2}{5}\0{3}{5}\0{4}{5}\0{5}{5}
\0{0}{4}\0{1}{4}\0{2}{4}\B{3}{4}\0{4}{4}\0{5}{4}
\0{0}{3}\1{1}{3}\0{2}{3}\1{3}{3}\A{4}{3}\0{5}{3}
\0{0}{2}\A{1}{2}\1{2}{2}\0{3}{2}\1{4}{2}\0{5}{2}
\0{0}{1}\0{1}{1}\B{2}{1}\0{3}{1}\0{4}{1}\0{5}{1}
\0{0}{0}\0{1}{0}\0{2}{0}\0{3}{0}\0{4}{0}\0{5}{0}
\mygrid{6}{6}
\end{picture}}}
\leftrightarrow
\vcenter{\hbox{\begin{picture}(6,6)
\0{0}{5}\0{1}{5}\0{2}{5}\0{3}{5}\0{4}{5}\0{5}{5}
\0{0}{4}\0{1}{4}\B{2}{4}\0{3}{4}\0{4}{4}\0{5}{4}
\0{0}{3}\A{1}{3}\0{2}{3}\1{3}{3}\1{4}{3}\0{5}{3}
\0{0}{2}\1{1}{2}\1{2}{2}\0{3}{2}\A{4}{2}\0{5}{2}
\0{0}{1}\0{1}{1}\0{2}{1}\B{3}{1}\0{4}{1}\0{5}{1}
\0{0}{0}\0{1}{0}\0{2}{0}\0{3}{0}\0{4}{0}\0{5}{0}
\mygrid{6}{6}
\end{picture}}}
\smallskip\\
\text{(a) `clock' ($a=0$), `block' ($a=1$)}
&
\text{(b) `clock' ($a=0$), `snake' ($a=1$)}
\medskip\\
\vcenter{\hbox{\begin{picture}(6,6)
\0{0}{5}\0{1}{5}\0{2}{5}\0{3}{5}\0{4}{5}\0{5}{5}
\0{0}{4}\0{1}{4}\A{2}{4}\B{3}{4}\0{4}{4}\0{5}{4}
\0{0}{3}\1{1}{3}\0{2}{3}\1{3}{3}\0{4}{3}\0{5}{3}
\0{0}{2}\0{1}{2}\1{2}{2}\0{3}{2}\1{4}{2}\0{5}{2}
\0{0}{1}\0{1}{1}\B{2}{1}\A{3}{1}\0{4}{1}\0{5}{1}
\0{0}{0}\0{1}{0}\0{2}{0}\0{3}{0}\0{4}{0}\0{5}{0}
\mygrid{6}{6}
\end{picture}}}
\leftrightarrow
\vcenter{\hbox{\begin{picture}(6,6)
\0{0}{5}\0{1}{5}\0{2}{5}\0{3}{5}\0{4}{5}\0{5}{5}
\0{0}{4}\0{1}{4}\1{2}{4}\0{3}{4}\0{4}{4}\0{5}{4}
\0{0}{3}\A{1}{3}\0{2}{3}\1{3}{3}\B{4}{3}\0{5}{3}
\0{0}{2}\B{1}{2}\1{2}{2}\0{3}{2}\A{4}{2}\0{5}{2}
\0{0}{1}\0{1}{1}\0{2}{1}\1{3}{1}\0{4}{1}\0{5}{1}
\0{0}{0}\0{1}{0}\0{2}{0}\0{3}{0}\0{4}{0}\0{5}{0}
\mygrid{6}{6}
\end{picture}}}
&
\vcenter{\hbox{\begin{picture}(6,6)
\0{0}{5}\0{1}{5}\0{2}{5}\0{3}{5}\0{4}{5}\0{5}{5}
\0{0}{4}\0{1}{4}\A{2}{4}\1{3}{4}\0{4}{4}\0{5}{4}
\0{0}{3}\1{1}{3}\0{2}{3}\B{3}{3}\A{4}{3}\0{5}{3}
\0{0}{2}\A{1}{2}\B{2}{2}\0{3}{2}\1{4}{2}\0{5}{2}
\0{0}{1}\0{1}{1}\1{2}{1}\A{3}{1}\0{4}{1}\0{5}{1}
\0{0}{0}\0{1}{0}\0{2}{0}\0{3}{0}\0{4}{0}\0{5}{0}
\mygrid{6}{6}
\end{picture}}}
\leftrightarrow
\vcenter{\hbox{\begin{picture}(6,6)
\0{0}{5}\0{1}{5}\0{2}{5}\0{3}{5}\0{4}{5}\0{5}{5}
\0{0}{4}\0{1}{4}\1{2}{4}\A{3}{4}\0{4}{4}\0{5}{4}
\0{0}{3}\A{1}{3}\0{2}{3}\B{3}{3}\1{4}{3}\0{5}{3}
\0{0}{2}\1{1}{2}\B{2}{2}\0{3}{2}\A{4}{2}\0{5}{2}
\0{0}{1}\0{1}{1}\A{2}{1}\1{3}{1}\0{4}{1}\0{5}{1}
\0{0}{0}\0{1}{0}\0{2}{0}\0{3}{0}\0{4}{0}\0{5}{0}
\mygrid{6}{6}
\end{picture}}}
\smallskip\\
\text{(c) `clock' ($a=0$), `barge' ($a=1$)}
&
\text{(d) `clock' ($a=0$), `pond' ($a=1$)}
\medskip\\
\vcenter{\hbox{\begin{picture}(6,6)
\0{0}{5}\0{1}{5}\0{2}{5}\0{3}{5}\0{4}{5}\0{5}{5}
\0{0}{4}\0{1}{4}\0{2}{4}\1{3}{4}\0{4}{4}\0{5}{4}
\0{0}{3}\1{1}{3}\A{2}{3}\B{3}{3}\0{4}{3}\0{5}{3}
\0{0}{2}\0{1}{2}\B{2}{2}\A{3}{2}\1{4}{2}\0{5}{2}
\0{0}{1}\0{1}{1}\1{2}{1}\0{3}{1}\0{4}{1}\0{5}{1}
\0{0}{0}\0{1}{0}\0{2}{0}\0{3}{0}\0{4}{0}\0{5}{0}
\mygrid{6}{6}
\end{picture}}}
\leftrightarrow
\vcenter{\hbox{\begin{picture}(6,6)
\0{0}{5}\0{1}{5}\0{2}{5}\0{3}{5}\0{4}{5}\0{5}{5}
\0{0}{4}\0{1}{4}\1{2}{4}\0{3}{4}\0{4}{4}\0{5}{4}
\0{0}{3}\0{1}{3}\A{2}{3}\B{3}{3}\1{4}{3}\0{5}{3}
\0{0}{2}\1{1}{2}\B{2}{2}\A{3}{2}\0{4}{2}\0{5}{2}
\0{0}{1}\0{1}{1}\0{2}{1}\1{3}{1}\0{4}{1}\0{5}{1}
\0{0}{0}\0{1}{0}\0{2}{0}\0{3}{0}\0{4}{0}\0{5}{0}
\mygrid{6}{6}
\end{picture}}}
\smallskip\\
\text{(e) `clock' ($a=0$), `clock' ($a=1$)}
\end{array}$
\end{center}
\caption{Solutions to equation (\ref{maxlife}) with period 2 giving `clock' for $a=0$.  For $a=1$, they give a static solution ((a)--(d)) or `clock' rotated by 90 degree ((e)).}
\label{fig:max clock}
\end{figure}
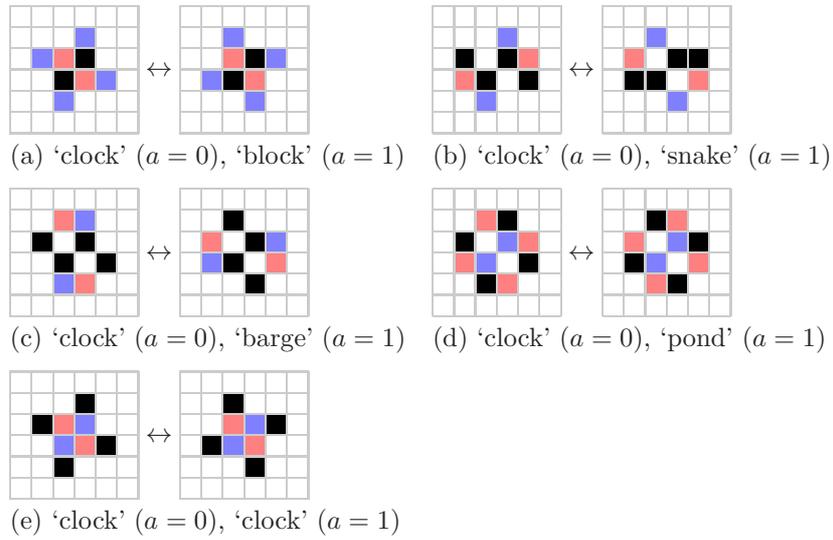

\begin{figure}[hbtp]
\begin{center}
$\begin{array}{ccc}
\multicolumn{3}{c}{\vcenter{\hbox{\begin{picture}(6,6)
\0{0}{5}\0{1}{5}\0{2}{5}\0{3}{5}\0{4}{5}\0{5}{5}
\0{0}{4}\0{1}{4}\0{2}{4}\1{3}{4}\0{4}{4}\0{5}{4}
\0{0}{3}\1{1}{3}\0{2}{3}\1{3}{3}\0{4}{3}\0{5}{3}
\0{0}{2}\0{1}{2}\1{2}{2}\0{3}{2}\1{4}{2}\0{5}{2}
\0{0}{1}\0{1}{1}\1{2}{1}\0{3}{1}\0{4}{1}\0{5}{1}
\0{0}{0}\0{1}{0}\0{2}{0}\0{3}{0}\0{4}{0}\0{5}{0}
\mygrid{6}{6}
\end{picture}}}
\leftrightarrow
\vcenter{\hbox{\begin{picture}(6,6)
\0{0}{5}\0{1}{5}\0{2}{5}\0{3}{5}\0{4}{5}\0{5}{5}
\0{0}{4}\0{1}{4}\1{2}{4}\0{3}{4}\0{4}{4}\0{5}{4}
\0{0}{3}\0{1}{3}\0{2}{3}\1{3}{3}\1{4}{3}\0{5}{3}
\0{0}{2}\1{1}{2}\1{2}{2}\0{3}{2}\0{4}{2}\0{5}{2}
\0{0}{1}\0{1}{1}\0{2}{1}\1{3}{1}\0{4}{1}\0{5}{1}
\0{0}{0}\0{1}{0}\0{2}{0}\0{3}{0}\0{4}{0}\0{5}{0}
\mygrid{6}{6}
\end{picture}}}}
\smallskip\\
\multicolumn{3}{c}{\text{`clock'}}
\end{array}$
\medskip\\
$\begin{array}{ccccc}
\begin{picture}(6,6)
\0{0}{5}\0{1}{5}\0{2}{5}\0{3}{5}\0{4}{5}\0{5}{5}
\0{0}{4}\0{1}{4}\0{2}{4}\0{3}{4}\0{4}{4}\0{5}{4}
\0{0}{3}\0{1}{3}\1{2}{3}\1{3}{3}\0{4}{3}\0{5}{3}
\0{0}{2}\0{1}{2}\1{2}{2}\1{3}{2}\0{4}{2}\0{5}{2}
\0{0}{1}\0{1}{1}\0{2}{1}\0{3}{1}\0{4}{1}\0{5}{1}
\0{0}{0}\0{1}{0}\0{2}{0}\0{3}{0}\0{4}{0}\0{5}{0}
\mygrid{6}{6}
\end{picture}
&
\begin{picture}(6,6)
\0{0}{5}\0{1}{5}\0{2}{5}\0{3}{5}\0{4}{5}\0{5}{5}
\0{0}{4}\0{1}{4}\0{2}{4}\0{3}{4}\0{4}{4}\0{5}{4}
\0{0}{3}\1{1}{3}\0{2}{3}\1{3}{3}\1{4}{3}\0{5}{3}
\0{0}{2}\1{1}{2}\1{2}{2}\0{3}{2}\1{4}{2}\0{5}{2}
\0{0}{1}\0{1}{1}\0{2}{1}\0{3}{1}\0{4}{1}\0{5}{1}
\0{0}{0}\0{1}{0}\0{2}{0}\0{3}{0}\0{4}{0}\0{5}{0}
\mygrid{6}{6}
\end{picture}
&
\begin{picture}(6,6)
\0{0}{5}\0{1}{5}\0{2}{5}\0{3}{5}\0{4}{5}\0{5}{5}
\0{0}{4}\0{1}{4}\1{2}{4}\0{3}{4}\0{4}{4}\0{5}{4}
\0{0}{3}\1{1}{3}\0{2}{3}\1{3}{3}\0{4}{3}\0{5}{3}
\0{0}{2}\0{1}{2}\1{2}{2}\0{3}{2}\1{4}{2}\0{5}{2}
\0{0}{1}\0{1}{1}\0{2}{1}\1{3}{1}\0{4}{1}\0{5}{1}
\0{0}{0}\0{1}{0}\0{2}{0}\0{3}{0}\0{4}{0}\0{5}{0}
\mygrid{6}{6}
\end{picture}
&
\begin{picture}(6,6)
\0{0}{5}\0{1}{5}\0{2}{5}\0{3}{5}\0{4}{5}\0{5}{5}
\0{0}{4}\0{1}{4}\1{2}{4}\1{3}{4}\0{4}{4}\0{5}{4}
\0{0}{3}\1{1}{3}\0{2}{3}\0{3}{3}\1{4}{3}\0{5}{3}
\0{0}{2}\1{1}{2}\0{2}{2}\0{3}{2}\1{4}{2}\0{5}{2}
\0{0}{1}\0{1}{1}\1{2}{1}\1{3}{1}\0{4}{1}\0{5}{1}
\0{0}{0}\0{1}{0}\0{2}{0}\0{3}{0}\0{4}{0}\0{5}{0}
\mygrid{6}{6}
\end{picture}
\smallskip\\
\text{`block'}
&
\text{`snake'}
&
\text{`barge'}
&
\text{`pond'}
\end{array}$
\end{center}
\caption{Solutions to GoL included in Figure~\ref{fig:max clock}}
\label{fig:clock}
\end{figure}
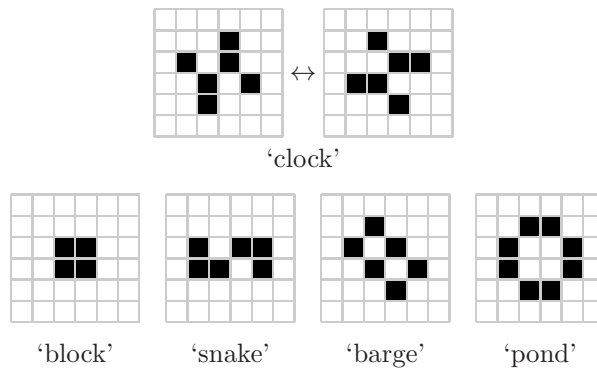

%%%%%%%%%%%%%%%%%%%%%%%%%%%%%%%%%%%%%%%%%%%%%%%%%%
  The next examples shown in Figure~\ref{fig:max clock} are 5 types of solutions giving the `clock' and another of GoL in a special case.
All solutions give `clock' for $a=0$. For $a=1$, static solution of GoL or another `clock' rotated by 90 degree are given.  The solutions of GoL described here are shown in Figure~\ref{fig:clock}.  Note that `clock' is periodic with period 2 and the other solutions are all static.
%%%%%%%%%%%%%%%%%%%%%%%%%%%%%%%%%%%%%%%%%%%%%%%%%%
\subsection{`Toad' type of solution}  \label{subsec:toad}
%%%%%%%%%%%%%%%%%%%%%%%%%%%%%%%%%%%%%%%%%%%%%%%%%%
  The next group of solutions include `toad' of GoL for $a=0$.  Figure~\ref{fig:max toad} shows the solutions and Figure~\ref{fig:toad} shows `toad'.
\begin{figure}[hbtp]
\begin{center}
$\begin{array}{ll}
\vcenter{\hbox{\begin{picture}(6,6)
\0{0}{5}\0{1}{5}\0{2}{5}\0{3}{5}\0{4}{5}\0{5}{5}
\0{0}{4}\0{1}{4}\0{2}{4}\B{3}{4}\0{4}{4}\0{5}{4}
\0{0}{3}\0{1}{3}\1{2}{3}\1{3}{3}\0{4}{3}\0{5}{3}
\0{0}{2}\0{1}{2}\1{2}{2}\1{3}{2}\0{4}{2}\0{5}{2}
\0{0}{1}\0{1}{1}\B{2}{1}\0{3}{1}\0{4}{1}\0{5}{1}
\0{0}{0}\0{1}{0}\0{2}{0}\0{3}{0}\0{4}{0}\0{5}{0}
\mygrid{6}{6}
\end{picture}}}
\leftrightarrow
\vcenter{\hbox{\begin{picture}(6,6)
\0{0}{5}\0{1}{5}\0{2}{5}\0{3}{5}\0{4}{5}\0{5}{5}
\0{0}{4}\0{1}{4}\B{2}{4}\B{3}{4}\0{4}{4}\0{5}{4}
\0{0}{3}\0{1}{3}\A{2}{3}\A{3}{3}\B{4}{3}\0{5}{3}
\0{0}{2}\B{1}{2}\A{2}{2}\A{3}{2}\0{4}{2}\0{5}{2}
\0{0}{1}\0{1}{1}\B{2}{1}\B{3}{1}\0{4}{1}\0{5}{1}
\0{0}{0}\0{1}{0}\0{2}{0}\0{3}{0}\0{4}{0}\0{5}{0}
\mygrid{6}{6}
\end{picture}}}
&
\vcenter{\hbox{\begin{picture}(6,6)
\0{0}{5}\0{1}{5}\0{2}{5}\0{3}{5}\0{4}{5}\0{5}{5}
\0{0}{4}\0{1}{4}\0{2}{4}\1{3}{4}\0{4}{4}\0{5}{4}
\0{0}{3}\0{1}{3}\1{2}{3}\B{3}{3}\A{4}{3}\0{5}{3}
\0{0}{2}\A{1}{2}\B{2}{2}\1{3}{2}\0{4}{2}\0{5}{2}
\0{0}{1}\0{1}{1}\1{2}{1}\0{3}{1}\0{4}{1}\0{5}{1}
\0{0}{0}\0{1}{0}\0{2}{0}\0{3}{0}\0{4}{0}\0{5}{0}
\mygrid{6}{6}
\end{picture}}}
\leftrightarrow
\vcenter{\hbox{\begin{picture}(6,6)
\0{0}{5}\0{1}{5}\0{2}{5}\0{3}{5}\0{4}{5}\0{5}{5}
\0{0}{4}\0{1}{4}\B{2}{4}\1{3}{4}\0{4}{4}\0{5}{4}
\0{0}{3}\0{1}{3}\A{2}{3}\0{3}{3}\1{4}{3}\0{5}{3}
\0{0}{2}\1{1}{2}\0{2}{2}\A{3}{2}\0{4}{2}\0{5}{2}
\0{0}{1}\0{1}{1}\1{2}{1}\B{3}{1}\0{4}{1}\0{5}{1}
\0{0}{0}\0{1}{0}\0{2}{0}\0{3}{0}\0{4}{0}\0{5}{0}
\mygrid{6}{6}
\end{picture}}}
\smallskip\\
\text{(a) `toad' ($a=0$), `block' ($a=1$)}
&
\text{(b) `toad' ($a=0$), `barge' ($a=1$)}
\medskip\\
\vcenter{\hbox{\begin{picture}(6,6)
\0{0}{5}\0{1}{5}\0{2}{5}\0{3}{5}\0{4}{5}\0{5}{5}
\0{0}{4}\0{1}{4}\A{2}{4}\1{3}{4}\0{4}{4}\0{5}{4}
\0{0}{3}\A{1}{3}\B{2}{3}\B{3}{3}\A{4}{3}\0{5}{3}
\0{0}{2}\A{1}{2}\B{2}{2}\B{3}{2}\A{4}{2}\0{5}{2}
\0{0}{1}\0{1}{1}\1{2}{1}\A{3}{1}\0{4}{1}\0{5}{1}
\0{0}{0}\0{1}{0}\0{2}{0}\0{3}{0}\0{4}{0}\0{5}{0}
\mygrid{6}{6}
\end{picture}}}
\leftrightarrow
\vcenter{\hbox{\begin{picture}(6,6)
\0{0}{5}\0{1}{5}\0{2}{5}\0{3}{5}\0{4}{5}\0{5}{5}
\0{0}{4}\0{1}{4}\1{2}{4}\1{3}{4}\0{4}{4}\0{5}{4}
\0{0}{3}\A{1}{3}\0{2}{3}\0{3}{3}\1{4}{3}\0{5}{3}
\0{0}{2}\1{1}{2}\0{2}{2}\0{3}{2}\A{4}{2}\0{5}{2}
\0{0}{1}\0{1}{1}\1{2}{1}\1{3}{1}\0{4}{1}\0{5}{1}
\0{0}{0}\0{1}{0}\0{2}{0}\0{3}{0}\0{4}{0}\0{5}{0}
\mygrid{6}{6}
\end{picture}}}
&
\vcenter{\hbox{\begin{picture}(6,6)
\0{0}{5}\0{1}{5}\0{2}{5}\0{3}{5}\0{4}{5}\0{5}{5}
\0{0}{4}\0{1}{4}\A{2}{4}\1{3}{4}\0{4}{4}\0{5}{4}
\0{0}{3}\0{1}{3}\1{2}{3}\B{3}{3}\0{4}{3}\0{5}{3}
\0{0}{2}\0{1}{2}\B{2}{2}\1{3}{2}\0{4}{2}\0{5}{2}
\0{0}{1}\0{1}{1}\1{2}{1}\A{3}{1}\0{4}{1}\0{5}{1}
\0{0}{0}\0{1}{0}\0{2}{0}\0{3}{0}\0{4}{0}\0{5}{0}
\mygrid{6}{6}
\end{picture}}}
\leftrightarrow
\vcenter{\hbox{\begin{picture}(6,6)
\0{0}{5}\0{1}{5}\0{2}{5}\0{3}{5}\0{4}{5}\0{5}{5}
\0{0}{4}\0{1}{4}\1{2}{4}\1{3}{4}\0{4}{4}\0{5}{4}
\0{0}{3}\0{1}{3}\A{2}{3}\0{3}{3}\B{4}{3}\0{5}{3}
\0{0}{2}\B{1}{2}\0{2}{2}\A{3}{2}\0{4}{2}\0{5}{2}
\0{0}{1}\0{1}{1}\1{2}{1}\1{3}{1}\0{4}{1}\0{5}{1}
\0{0}{0}\0{1}{0}\0{2}{0}\0{3}{0}\0{4}{0}\0{5}{0}
\mygrid{6}{6}
\end{picture}}}
\smallskip\\
\text{(c) `toad' ($a=0$), `pond' ($a=1$)}
&
\text{(d) `toad' ($a=0$), `snake' ($a=1$)}
\medskip\\
\vcenter{\hbox{\begin{picture}(6,6)
\0{0}{5}\0{1}{5}\0{2}{5}\0{3}{5}\0{4}{5}\0{5}{5}
\0{0}{4}\0{1}{4}\A{2}{4}\1{3}{4}\0{4}{4}\0{5}{4}
\0{0}{3}\0{1}{3}\B{2}{3}\1{3}{3}\0{4}{3}\0{5}{3}
\0{0}{2}\0{1}{2}\1{2}{2}\B{3}{2}\0{4}{2}\0{5}{2}
\0{0}{1}\0{1}{1}\1{2}{1}\A{3}{1}\0{4}{1}\0{5}{1}
\0{0}{0}\0{1}{0}\0{2}{0}\0{3}{0}\0{4}{0}\0{5}{0}
\mygrid{6}{6}
\end{picture}}}
\leftrightarrow
\vcenter{\hbox{\begin{picture}(6,6)
\0{0}{5}\0{1}{5}\0{2}{5}\0{3}{5}\0{4}{5}\0{5}{5}
\0{0}{4}\0{1}{4}\1{2}{4}\1{3}{4}\0{4}{4}\0{5}{4}
\0{0}{3}\0{1}{3}\0{2}{3}\A{3}{3}\B{4}{3}\0{5}{3}
\0{0}{2}\B{1}{2}\A{2}{2}\0{3}{2}\0{4}{2}\0{5}{2}
\0{0}{1}\0{1}{1}\1{2}{1}\1{3}{1}\0{4}{1}\0{5}{1}
\0{0}{0}\0{1}{0}\0{2}{0}\0{3}{0}\0{4}{0}\0{5}{0}
\mygrid{6}{6}
\end{picture}}}
&
\vcenter{\hbox{\begin{picture}(6,6)
\0{0}{5}\0{1}{5}\0{2}{5}\0{3}{5}\0{4}{5}\0{5}{5}
\0{0}{4}\0{1}{4}\0{2}{4}\B{3}{4}\0{4}{4}\0{5}{4}
\0{0}{3}\A{1}{3}\1{2}{3}\B{3}{3}\A{4}{3}\0{5}{3}
\0{0}{2}\A{1}{2}\B{2}{2}\1{3}{2}\A{4}{2}\0{5}{2}
\0{0}{1}\0{1}{1}\B{2}{1}\0{3}{1}\0{4}{1}\0{5}{1}
\0{0}{0}\0{1}{0}\0{2}{0}\0{3}{0}\0{4}{0}\0{5}{0}
\mygrid{6}{6}
\end{picture}}}
\leftrightarrow
\vcenter{\hbox{\begin{picture}(6,6)
\0{0}{5}\0{1}{5}\0{2}{5}\0{3}{5}\0{4}{5}\0{5}{5}
\0{0}{4}\0{1}{4}\B{2}{4}\B{3}{4}\0{4}{4}\0{5}{4}
\0{0}{3}\A{1}{3}\A{2}{3}\0{3}{3}\1{4}{3}\0{5}{3}
\0{0}{2}\1{1}{2}\0{2}{2}\A{3}{2}\A{4}{2}\0{5}{2}
\0{0}{1}\0{1}{1}\B{2}{1}\B{3}{1}\0{4}{1}\0{5}{1}
\0{0}{0}\0{1}{0}\0{2}{0}\0{3}{0}\0{4}{0}\0{5}{0}
\mygrid{6}{6}
\end{picture}}}
\smallskip\\
\text{(e) `toad' ($a=0$), `snake' ($a=1$)}
&
\text{(f) `toad' ($a=0$), `snake' ($a=1$)}
\medskip\\
\vcenter{\hbox{\begin{picture}(6,6)
\0{0}{5}\0{1}{5}\0{2}{5}\0{3}{5}\0{4}{5}\0{5}{5}
\0{0}{4}\0{1}{4}\0{2}{4}\B{3}{4}\0{4}{4}\0{5}{4}
\0{0}{3}\A{1}{3}\B{2}{3}\1{3}{3}\A{4}{3}\0{5}{3}
\0{0}{2}\A{1}{2}\1{2}{2}\B{3}{2}\A{4}{2}\0{5}{2}
\0{0}{1}\0{1}{1}\B{2}{1}\0{3}{1}\0{4}{1}\0{5}{1}
\0{0}{0}\0{1}{0}\0{2}{0}\0{3}{0}\0{4}{0}\0{5}{0}
\mygrid{6}{6}
\end{picture}}}
\leftrightarrow
\vcenter{\hbox{\begin{picture}(6,6)
\0{0}{5}\0{1}{5}\0{2}{5}\0{3}{5}\0{4}{5}\0{5}{5}
\0{0}{4}\0{1}{4}\B{2}{4}\B{3}{4}\0{4}{4}\0{5}{4}
\0{0}{3}\A{1}{3}\0{2}{3}\A{3}{3}\1{4}{3}\0{5}{3}
\0{0}{2}\1{1}{2}\A{2}{2}\0{3}{2}\A{4}{2}\0{5}{2}
\0{0}{1}\0{1}{1}\B{2}{1}\B{3}{1}\0{4}{1}\0{5}{1}
\0{0}{0}\0{1}{0}\0{2}{0}\0{3}{0}\0{4}{0}\0{5}{0}
\mygrid{6}{6}
\end{picture}}}
&
\vcenter{\hbox{\begin{picture}(6,6)
\0{0}{5}\0{1}{5}\0{2}{5}\0{3}{5}\0{4}{5}\0{5}{5}
\0{0}{4}\0{1}{4}\0{2}{4}\1{3}{4}\0{4}{4}\0{5}{4}
\0{0}{3}\A{1}{3}\1{2}{3}\B{3}{3}\0{4}{3}\0{5}{3}
\0{0}{2}\0{1}{2}\B{2}{2}\1{3}{2}\A{4}{2}\0{5}{2}
\0{0}{1}\0{1}{1}\1{2}{1}\0{3}{1}\0{4}{1}\0{5}{1}
\0{0}{0}\0{1}{0}\0{2}{0}\0{3}{0}\0{4}{0}\0{5}{0}
\mygrid{6}{6}
\end{picture}}}
\leftrightarrow
\vcenter{\hbox{\begin{picture}(6,6)
\0{0}{5}\0{1}{5}\0{2}{5}\0{3}{5}\0{4}{5}\0{5}{5}
\0{0}{4}\0{1}{4}\1{2}{4}\B{3}{4}\0{4}{4}\0{5}{4}
\0{0}{3}\0{1}{3}\A{2}{3}\0{3}{3}\1{4}{3}\0{5}{3}
\0{0}{2}\1{1}{2}\0{2}{2}\A{3}{2}\0{4}{2}\0{5}{2}
\0{0}{1}\0{1}{1}\B{2}{1}\1{3}{1}\0{4}{1}\0{5}{1}
\0{0}{0}\0{1}{0}\0{2}{0}\0{3}{0}\0{4}{0}\0{5}{0}
\mygrid{6}{6}
\end{picture}}}
\smallskip\\
\text{(g) `toad' ($a=0$), `snake' ($a=1$)}
&
\text{(h) `toad' ($a=0$), `clock' ($a=1$)}
\medskip\\
\vcenter{\hbox{\begin{picture}(6,6)
\0{0}{5}\0{1}{5}\0{2}{5}\0{3}{5}\0{4}{5}\0{5}{5}
\0{0}{4}\0{1}{4}\0{2}{4}\1{3}{4}\0{4}{4}\0{5}{4}
\0{0}{3}\A{1}{3}\B{2}{3}\1{3}{3}\0{4}{3}\0{5}{3}
\0{0}{2}\0{1}{2}\1{2}{2}\B{3}{2}\A{4}{2}\0{5}{2}
\0{0}{1}\0{1}{1}\1{2}{1}\0{3}{1}\0{4}{1}\0{5}{1}
\0{0}{0}\0{1}{0}\0{2}{0}\0{3}{0}\0{4}{0}\0{5}{0}
\mygrid{6}{6}
\end{picture}}}
\leftrightarrow
\vcenter{\hbox{\begin{picture}(6,6)
\0{0}{5}\0{1}{5}\0{2}{5}\0{3}{5}\0{4}{5}\0{5}{5}
\0{0}{4}\0{1}{4}\1{2}{4}\B{3}{4}\0{4}{4}\0{5}{4}
\0{0}{3}\0{1}{3}\0{2}{3}\A{3}{3}\1{4}{3}\0{5}{3}
\0{0}{2}\1{1}{2}\A{2}{2}\0{3}{2}\0{4}{2}\0{5}{2}
\0{0}{1}\0{1}{1}\B{2}{1}\1{3}{1}\0{4}{1}\0{5}{1}
\0{0}{0}\0{1}{0}\0{2}{0}\0{3}{0}\0{4}{0}\0{5}{0}
\mygrid{6}{6}
\end{picture}}}
&
\vcenter{\hbox{\begin{picture}(6,6)
\0{0}{5}\0{1}{5}\0{2}{5}\0{3}{5}\0{4}{5}\0{5}{5}
\0{0}{4}\0{1}{4}\A{2}{4}\B{3}{4}\0{4}{4}\0{5}{4}
\0{0}{3}\0{1}{3}\1{2}{3}\1{3}{3}\0{4}{3}\0{5}{3}
\0{0}{2}\0{1}{2}\1{2}{2}\1{3}{2}\0{4}{2}\0{5}{2}
\0{0}{1}\0{1}{1}\B{2}{1}\A{3}{1}\0{4}{1}\0{5}{1}
\0{0}{0}\0{1}{0}\0{2}{0}\0{3}{0}\0{4}{0}\0{5}{0}
\mygrid{6}{6}
\end{picture}}}
\leftrightarrow
\vcenter{\hbox{\begin{picture}(6,6)
\0{0}{5}\0{1}{5}\0{2}{5}\0{3}{5}\0{4}{5}\0{5}{5}
\0{0}{4}\0{1}{4}\1{2}{4}\1{3}{4}\0{4}{4}\0{5}{4}
\0{0}{3}\A{1}{3}\0{2}{3}\0{3}{3}\B{4}{3}\0{5}{3}
\0{0}{2}\B{1}{2}\0{2}{2}\0{3}{2}\A{4}{2}\0{5}{2}
\0{0}{1}\0{1}{1}\1{2}{1}\1{3}{1}\0{4}{1}\0{5}{1}
\0{0}{0}\0{1}{0}\0{2}{0}\0{3}{0}\0{4}{0}\0{5}{0}
\mygrid{6}{6}
\end{picture}}}
\smallskip\\
\text{(i) `toad' ($a=0$), `clock' ($a=1$)}
&
\text{(j) `toad' ($a=0$), `toad' ($a=1$)}
\medskip\\
\vcenter{\hbox{\begin{picture}(6,6)
\0{0}{5}\0{1}{5}\0{2}{5}\0{3}{5}\0{4}{5}\0{5}{5}
\0{0}{4}\0{1}{4}\0{2}{4}\1{3}{4}\0{4}{4}\0{5}{4}
\0{0}{3}\A{1}{3}\B{2}{3}\B{3}{3}\A{4}{3}\0{5}{3}
\0{0}{2}\A{1}{2}\B{2}{2}\B{3}{2}\A{4}{2}\0{5}{2}
\0{0}{1}\0{1}{1}\1{2}{1}\0{3}{1}\0{4}{1}\0{5}{1}
\0{0}{0}\0{1}{0}\0{2}{0}\0{3}{0}\0{4}{0}\0{5}{0}
\mygrid{6}{6}
\end{picture}}}
\leftrightarrow
\vcenter{\hbox{\begin{picture}(6,6)
\0{0}{5}\0{1}{5}\0{2}{5}\0{3}{5}\0{4}{5}\0{5}{5}
\0{0}{4}\0{1}{4}\B{2}{4}\B{3}{4}\0{4}{4}\0{5}{4}
\0{0}{3}\0{1}{3}\A{2}{3}\A{3}{3}\1{4}{3}\0{5}{3}
\0{0}{2}\1{1}{2}\A{2}{2}\A{3}{2}\0{4}{2}\0{5}{2}
\0{0}{1}\0{1}{1}\B{2}{1}\B{3}{1}\0{4}{1}\0{5}{1}
\0{0}{0}\0{1}{0}\0{2}{0}\0{3}{0}\0{4}{0}\0{5}{0}
\mygrid{6}{6}
\end{picture}}}
&
\vcenter{\hbox{\begin{picture}(6,6)
\0{0}{5}\0{1}{5}\0{2}{5}\0{3}{5}\0{4}{5}\0{5}{5}
\0{0}{4}\0{1}{4}\0{2}{4}\B{3}{4}\0{4}{4}\0{5}{4}
\0{0}{3}\0{1}{3}\1{2}{3}\1{3}{3}\A{4}{3}\0{5}{3}
\0{0}{2}\A{1}{2}\1{2}{2}\1{3}{2}\0{4}{2}\0{5}{2}
\0{0}{1}\0{1}{1}\B{2}{1}\0{3}{1}\0{4}{1}\0{5}{1}
\0{0}{0}\0{1}{0}\0{2}{0}\0{3}{0}\0{4}{0}\0{5}{0}
\mygrid{6}{6}
\end{picture}}}
\leftrightarrow
\vcenter{\hbox{\begin{picture}(6,6)
\0{0}{5}\0{1}{5}\0{2}{5}\0{3}{5}\0{4}{5}\0{5}{5}
\0{0}{4}\0{1}{4}\B{2}{4}\1{3}{4}\0{4}{4}\0{5}{4}
\0{0}{3}\A{1}{3}\0{2}{3}\0{3}{3}\1{4}{3}\0{5}{3}
\0{0}{2}\1{1}{2}\0{2}{2}\0{3}{2}\A{4}{2}\0{5}{2}
\0{0}{1}\0{1}{1}\1{2}{1}\B{3}{1}\0{4}{1}\0{5}{1}
\0{0}{0}\0{1}{0}\0{2}{0}\0{3}{0}\0{4}{0}\0{5}{0}
\mygrid{6}{6}
\end{picture}}}
\smallskip\\
\text{(k) `toad' ($a=0$), `toad' ($a=1$)}
&
\text{(l) `toad' ($a=0$), `toad' ($a=1$)}
\medskip\\
\vcenter{\hbox{\begin{picture}(6,6)
\0{0}{5}\0{1}{5}\0{2}{5}\0{3}{5}\0{4}{5}\0{5}{5}
\0{0}{4}\0{1}{4}\0{2}{4}\B{3}{4}\0{4}{4}\0{5}{4}
\0{0}{3}\A{1}{3}\1{2}{3}\1{3}{3}\0{4}{3}\0{5}{3}
\0{0}{2}\0{1}{2}\1{2}{2}\1{3}{2}\A{4}{2}\0{5}{2}
\0{0}{1}\0{1}{1}\B{2}{1}\0{3}{1}\0{4}{1}\0{5}{1}
\0{0}{0}\0{1}{0}\0{2}{0}\0{3}{0}\0{4}{0}\0{5}{0}
\mygrid{6}{6}
\end{picture}}}
\leftrightarrow
\vcenter{\hbox{\begin{picture}(6,6)
\0{0}{5}\0{1}{5}\0{2}{5}\0{3}{5}\0{4}{5}\0{5}{5}
\0{0}{4}\0{1}{4}\1{2}{4}\B{3}{4}\0{4}{4}\0{5}{4}
\0{0}{3}\A{1}{3}\0{2}{3}\0{3}{3}\1{4}{3}\0{5}{3}
\0{0}{2}\1{1}{2}\0{2}{2}\0{3}{2}\A{4}{2}\0{5}{2}
\0{0}{1}\0{1}{1}\B{2}{1}\1{3}{1}\0{4}{1}\0{5}{1}
\0{0}{0}\0{1}{0}\0{2}{0}\0{3}{0}\0{4}{0}\0{5}{0}
\mygrid{6}{6}
\end{picture}}}
&
\vcenter{\hbox{\begin{picture}(6,6)
\0{0}{5}\0{1}{5}\0{2}{5}\0{3}{5}\0{4}{5}\0{5}{5}
\0{0}{4}\0{1}{4}\A{2}{4}\1{3}{4}\0{4}{4}\0{5}{4}
\0{0}{3}\A{1}{3}\B{2}{3}\B{3}{3}\0{4}{3}\0{5}{3}
\0{0}{2}\0{1}{2}\B{2}{2}\B{3}{2}\A{4}{2}\0{5}{2}
\0{0}{1}\0{1}{1}\1{2}{1}\A{3}{1}\0{4}{1}\0{5}{1}
\0{0}{0}\0{1}{0}\0{2}{0}\0{3}{0}\0{4}{0}\0{5}{0}
\mygrid{6}{6}
\end{picture}}}
\leftrightarrow
\vcenter{\hbox{\begin{picture}(6,6)
\0{0}{5}\0{1}{5}\0{2}{5}\0{3}{5}\0{4}{5}\0{5}{5}
\0{0}{4}\0{1}{4}\1{2}{4}\B{3}{4}\0{4}{4}\0{5}{4}
\0{0}{3}\0{1}{3}\A{2}{3}\A{3}{3}\B{4}{3}\0{5}{3}
\0{0}{2}\B{1}{2}\A{2}{2}\A{3}{2}\0{4}{2}\0{5}{2}
\0{0}{1}\0{1}{1}\B{2}{1}\1{3}{1}\0{4}{1}\0{5}{1}
\0{0}{0}\0{1}{0}\0{2}{0}\0{3}{0}\0{4}{0}\0{5}{0}
\mygrid{6}{6}
\end{picture}}}
\smallskip\\
\text{(m) `toad' ($a=0$), `toad' ($a=1$)}
&
\text{(n) `toad' ($a=0$), `toad' ($a=1$)}
\end{array}$
\end{center}
\caption{Solutions to equation (\ref{maxlife}) with period 2 giving `toad' for $a=0$.  For $a=1$, they give (a)-(g) a static solution, (h)-(i) `clock', and (j)-(n) another configuration of `toad'.}
\label{fig:max toad}
\end{figure}
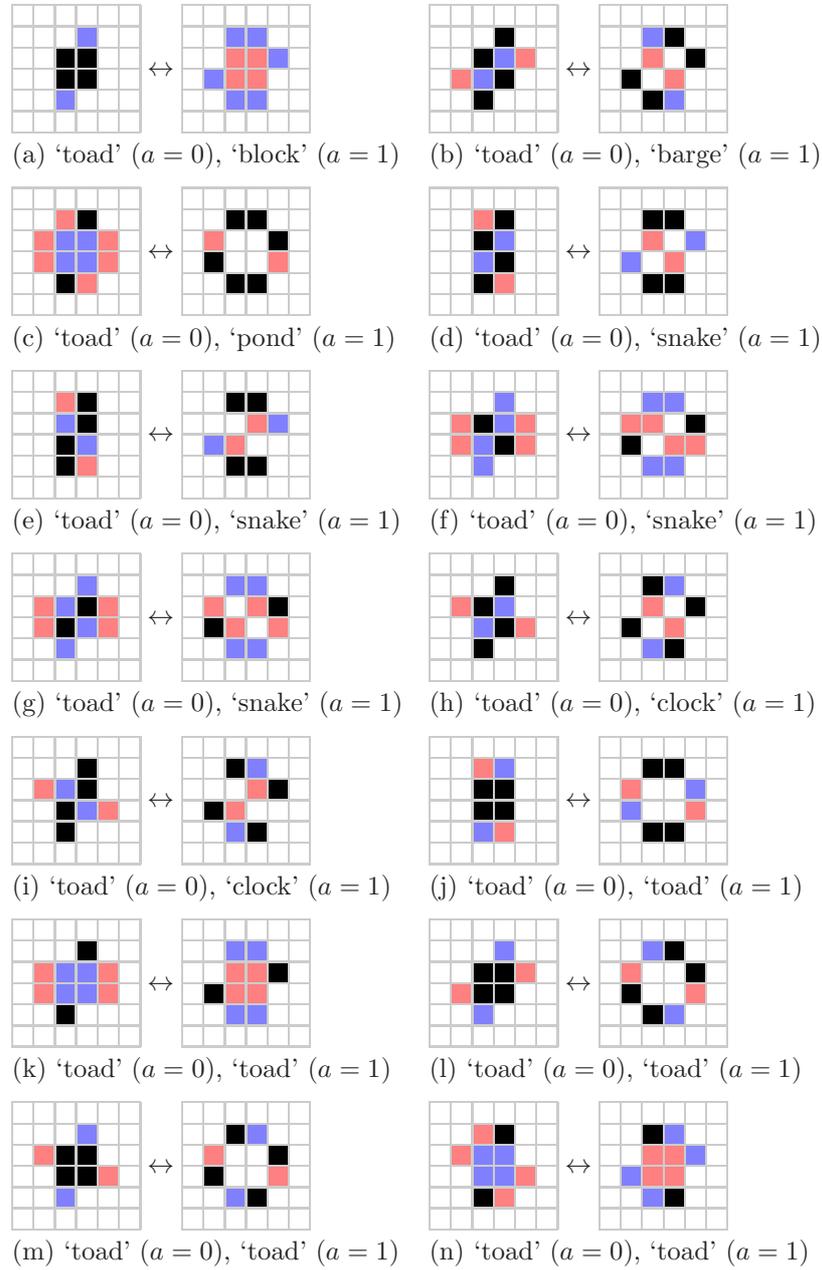
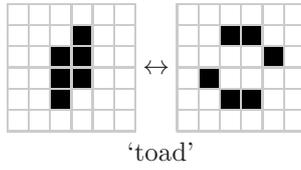
\begin{figure}[hbtp]
\begin{center}
$\vcenter{\hbox{\begin{picture}(6,6)
\0{0}{5}\0{1}{5}\0{2}{5}\0{3}{5}\0{4}{5}\0{5}{5}
\0{0}{4}\0{1}{4}\0{2}{4}\1{3}{4}\0{4}{4}\0{5}{4}
\0{0}{3}\0{1}{3}\1{2}{3}\1{3}{3}\0{4}{3}\0{5}{3}
\0{0}{2}\0{1}{2}\1{2}{2}\1{3}{2}\0{4}{2}\0{5}{2}
\0{0}{1}\0{1}{1}\1{2}{1}\0{3}{1}\0{4}{1}\0{5}{1}
\0{0}{0}\0{1}{0}\0{2}{0}\0{3}{0}\0{4}{0}\0{5}{0}
\mygrid{6}{6}
\end{picture}}}
\leftrightarrow
\vcenter{\hbox{\begin{picture}(6,6)
\0{0}{5}\0{1}{5}\0{2}{5}\0{3}{5}\0{4}{5}\0{5}{5}
\0{0}{4}\0{1}{4}\1{2}{4}\1{3}{4}\0{4}{4}\0{5}{4}
\0{0}{3}\0{1}{3}\0{2}{3}\0{3}{3}\1{4}{3}\0{5}{3}
\0{0}{2}\1{1}{2}\0{2}{2}\0{3}{2}\0{4}{2}\0{5}{2}
\0{0}{1}\0{1}{1}\1{2}{1}\1{3}{1}\0{4}{1}\0{5}{1}
\0{0}{0}\0{1}{0}\0{2}{0}\0{3}{0}\0{4}{0}\0{5}{0}
\mygrid{6}{6}
\end{picture}}}$
\smallskip\\
`toad'
\end{center}
\caption{`Toad' of GoL.}
\label{fig:toad}
\end{figure}
%%%%%%%%%%%%%%%%%%%%%%%%%%%%%%%%%%%%%%%%%%%%%%%%%%
\subsection{`Glider' type of solution}  \label{subsec:glider}
\begin{figure}[hbtp]
\begin{center}
$\vcenter{\hbox{\begin{picture}(6,6)
\0{0}{5}\0{1}{5}\0{2}{5}\0{3}{5}\0{4}{5}\0{5}{5}
\0{0}{4}\0{1}{4}\0{2}{4}\0{3}{4}\0{4}{4}\0{5}{4}
\0{0}{3}\0{1}{3}\1{2}{3}\1{3}{3}\0{4}{3}\0{5}{3}
\0{0}{2}\1{1}{2}\0{2}{2}\1{3}{2}\0{4}{2}\0{5}{2}
\0{0}{1}\0{1}{1}\0{2}{1}\1{3}{1}\0{4}{1}\0{5}{1}
\0{0}{0}\0{1}{0}\0{2}{0}\0{3}{0}\0{4}{0}\0{5}{0}
\mygrid{6}{6}
\end{picture}}}
\rightarrow
\vcenter{\hbox{\begin{picture}(6,6)
\0{0}{5}\0{1}{5}\0{2}{5}\0{3}{5}\0{4}{5}\0{5}{5}
\0{0}{4}\0{1}{4}\0{2}{4}\0{3}{4}\0{4}{4}\0{5}{4}
\0{0}{3}\0{1}{3}\1{2}{3}\1{3}{3}\0{4}{3}\0{5}{3}
\0{0}{2}\0{1}{2}\0{2}{2}\1{3}{2}\1{4}{2}\0{5}{2}
\0{0}{1}\0{1}{1}\1{2}{1}\0{3}{1}\0{4}{1}\0{5}{1}
\0{0}{0}\0{1}{0}\0{2}{0}\0{3}{0}\0{4}{0}\0{5}{0}
\mygrid{6}{6}
\end{picture}}}
\rightarrow
\vcenter{\hbox{\begin{picture}(6,6)
\0{0}{5}\0{1}{5}\0{2}{5}\0{3}{5}\0{4}{5}\0{5}{5}
\0{0}{4}\0{1}{4}\0{2}{4}\0{3}{4}\0{4}{4}\0{5}{4}
\0{0}{3}\0{1}{3}\1{2}{3}\1{3}{3}\1{4}{3}\0{5}{3}
\0{0}{2}\0{1}{2}\0{2}{2}\0{3}{2}\1{4}{2}\0{5}{2}
\0{0}{1}\0{1}{1}\0{2}{1}\1{3}{1}\0{4}{1}\0{5}{1}
\0{0}{0}\0{1}{0}\0{2}{0}\0{3}{0}\0{4}{0}\0{5}{0}
\mygrid{6}{6}
\end{picture}}}
\rightarrow
\vcenter{\hbox{\begin{picture}(6,6)
\0{0}{5}\0{1}{5}\0{2}{5}\0{3}{5}\0{4}{5}\0{5}{5}
\0{0}{4}\0{1}{4}\0{2}{4}\1{3}{4}\0{4}{4}\0{5}{4}
\0{0}{3}\0{1}{3}\0{2}{3}\1{3}{3}\1{4}{3}\0{5}{3}
\0{0}{2}\0{1}{2}\1{2}{2}\0{3}{2}\1{4}{2}\0{5}{2}
\0{0}{1}\0{1}{1}\0{2}{1}\0{3}{1}\0{4}{1}\0{5}{1}
\0{0}{0}\0{1}{0}\0{2}{0}\0{3}{0}\0{4}{0}\0{5}{0}
\mygrid{6}{6}
\end{picture}}}
\rightarrow
\vcenter{\hbox{\begin{picture}(6,6)
\0{0}{5}\0{1}{5}\0{2}{5}\0{3}{5}\0{4}{5}\0{5}{5}
\0{0}{4}\0{1}{4}\0{2}{4}\1{3}{4}\1{4}{4}\0{5}{4}
\0{0}{3}\0{1}{3}\1{2}{3}\0{3}{3}\1{4}{3}\0{5}{3}
\0{0}{2}\0{1}{2}\0{2}{2}\0{3}{2}\1{4}{2}\0{5}{2}
\0{0}{1}\0{1}{1}\0{2}{1}\0{3}{1}\0{4}{1}\0{5}{1}
\0{0}{0}\0{1}{0}\0{2}{0}\0{3}{0}\0{4}{0}\0{5}{0}
\mygrid{6}{6}
\end{picture}}}$
\caption{`Glider' of GoL.}
\label{fig:glider}
\end{center}
\end{figure}
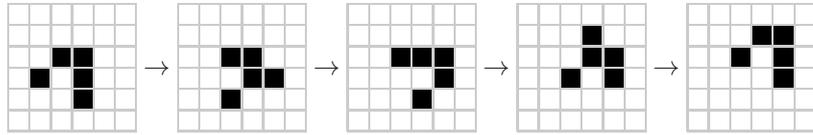

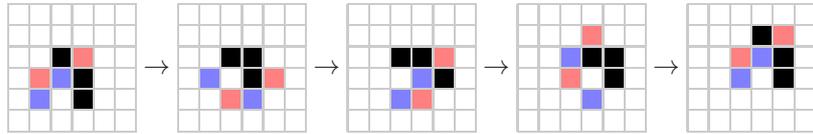
\begin{figure}[hbtp]
\begin{center}
$\vcenter{\hbox{\begin{picture}(6,6)
\0{0}{5}\0{1}{5}\0{2}{5}\0{3}{5}\0{4}{5}\0{5}{5}
\0{0}{4}\0{1}{4}\0{2}{4}\0{3}{4}\0{4}{4}\0{5}{4}
\0{0}{3}\0{1}{3}\1{2}{3}\A{3}{3}\0{4}{3}\0{5}{3}
\0{0}{2}\A{1}{2}\B{2}{2}\1{3}{2}\0{4}{2}\0{5}{2}
\0{0}{1}\B{1}{1}\0{2}{1}\1{3}{1}\0{4}{1}\0{5}{1}
\0{0}{0}\0{1}{0}\0{2}{0}\0{3}{0}\0{4}{0}\0{5}{0}
\mygrid{6}{6}
\end{picture}}}
\rightarrow
\vcenter{\hbox{\begin{picture}(6,6)
\0{0}{5}\0{1}{5}\0{2}{5}\0{3}{5}\0{4}{5}\0{5}{5}
\0{0}{4}\0{1}{4}\0{2}{4}\0{3}{4}\0{4}{4}\0{5}{4}
\0{0}{3}\0{1}{3}\1{2}{3}\1{3}{3}\0{4}{3}\0{5}{3}
\0{0}{2}\B{1}{2}\0{2}{2}\1{3}{2}\A{4}{2}\0{5}{2}
\0{0}{1}\0{1}{1}\A{2}{1}\B{3}{1}\0{4}{1}\0{5}{1}
\0{0}{0}\0{1}{0}\0{2}{0}\0{3}{0}\0{4}{0}\0{5}{0}
\mygrid{6}{6}
\end{picture}}}
\rightarrow
\vcenter{\hbox{\begin{picture}(6,6)
\0{0}{5}\0{1}{5}\0{2}{5}\0{3}{5}\0{4}{5}\0{5}{5}
\0{0}{4}\0{1}{4}\0{2}{4}\0{3}{4}\0{4}{4}\0{5}{4}
\0{0}{3}\0{1}{3}\1{2}{3}\1{3}{3}\A{4}{3}\0{5}{3}
\0{0}{2}\0{1}{2}\0{2}{2}\B{3}{2}\1{4}{2}\0{5}{2}
\0{0}{1}\0{1}{1}\B{2}{1}\A{3}{1}\0{4}{1}\0{5}{1}
\0{0}{0}\0{1}{0}\0{2}{0}\0{3}{0}\0{4}{0}\0{5}{0}
\mygrid{6}{6}
\end{picture}}}
\rightarrow
\vcenter{\hbox{\begin{picture}(6,6)
\0{0}{5}\0{1}{5}\0{2}{5}\0{3}{5}\0{4}{5}\0{5}{5}
\0{0}{4}\0{1}{4}\0{2}{4}\A{3}{4}\0{4}{4}\0{5}{4}
\0{0}{3}\0{1}{3}\B{2}{3}\1{3}{3}\1{4}{3}\0{5}{3}
\0{0}{2}\0{1}{2}\A{2}{2}\0{3}{2}\1{4}{2}\0{5}{2}
\0{0}{1}\0{1}{1}\0{2}{1}\B{3}{1}\0{4}{1}\0{5}{1}
\0{0}{0}\0{1}{0}\0{2}{0}\0{3}{0}\0{4}{0}\0{5}{0}
\mygrid{6}{6}
\end{picture}}}
\rightarrow
\vcenter{\hbox{\begin{picture}(6,6)
\0{0}{5}\0{1}{5}\0{2}{5}\0{3}{5}\0{4}{5}\0{5}{5}
\0{0}{4}\0{1}{4}\0{2}{4}\1{3}{4}\A{4}{4}\0{5}{4}
\0{0}{3}\0{1}{3}\A{2}{3}\B{3}{3}\1{4}{3}\0{5}{3}
\0{0}{2}\0{1}{2}\B{2}{2}\0{3}{2}\1{4}{2}\0{5}{2}
\0{0}{1}\0{1}{1}\0{2}{1}\0{3}{1}\0{4}{1}\0{5}{1}
\0{0}{0}\0{1}{0}\0{2}{0}\0{3}{0}\0{4}{0}\0{5}{0}
\mygrid{6}{6}
\end{picture}}}$
\caption{Solution to equation (\ref{maxlife}) giving `glider' with different time phase for $a=0$ and 1.}
\label{fig:max glider}
\end{center}
\end{figure}

%%%%%%%%%%%%%%%%%%%%%%%%%%%%%%%%%%%%%%%%%%%%%%%%%%
  There are solutions giving a moving pattern of GoL.  One of the simplest solutions is called `glider' shown in Figure~\ref{fig:glider}.
Figure~\ref{fig:max glider} shows the real-valued solution and it coincides with that of Figure~\ref{fig:glider} if $a=1$ and gives another glider of different time phase if $a=0$.
%%%%%%%%%%%%%%%%%%%%%%%%%%%%%%%%%%%%%%%%%%%%%%%%%%
\subsection{More general solution}
\begin{figure}[hbtp]
\begin{center}
$\begin{array}{cc}
\vcenter{\hbox{\includegraphics[scale=0.25]{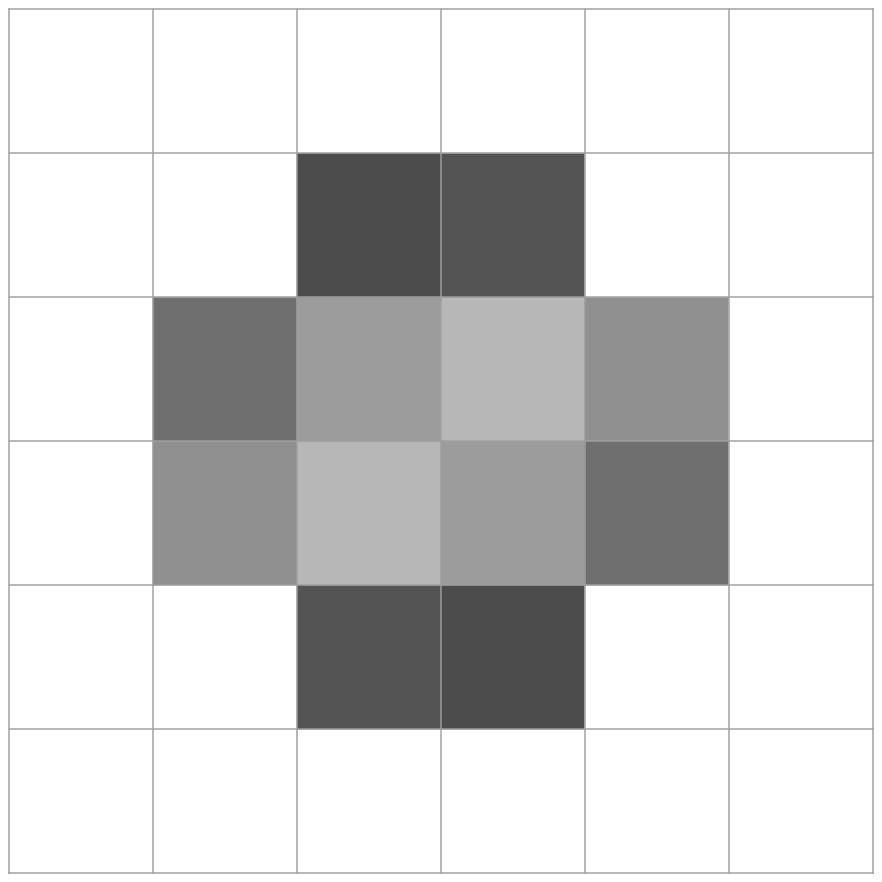}}}
\leftrightarrow
\vcenter{\hbox{\includegraphics[scale=0.25]{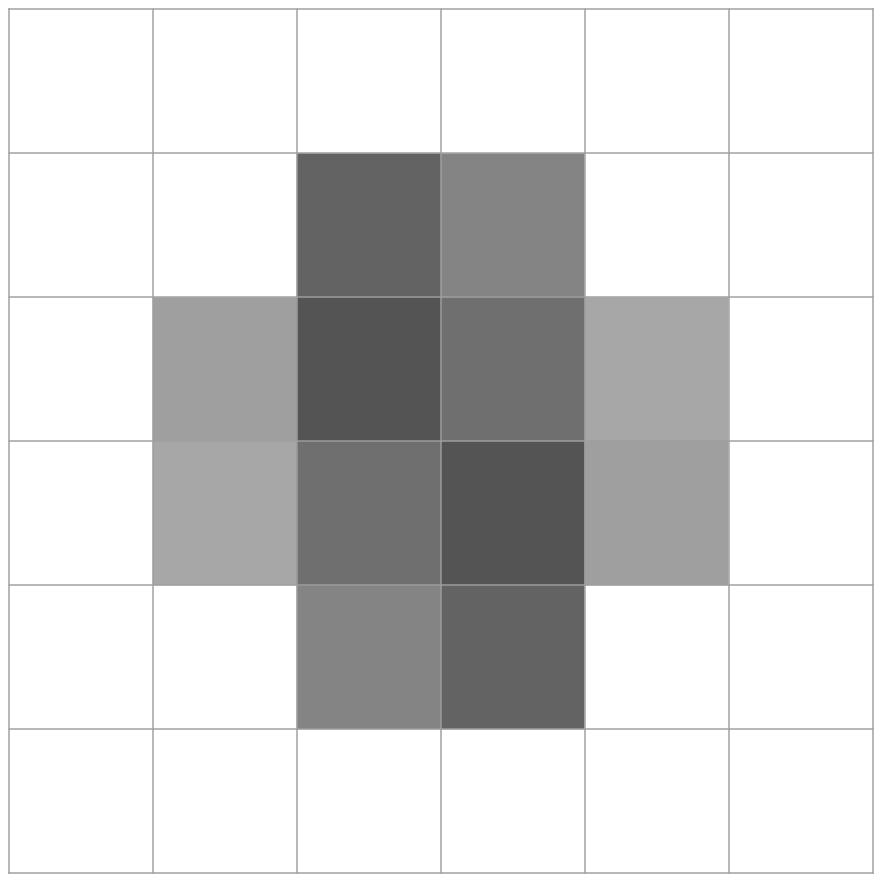}}}
&
\qquad
\vcenter{\hbox{\includegraphics[scale=0.25]{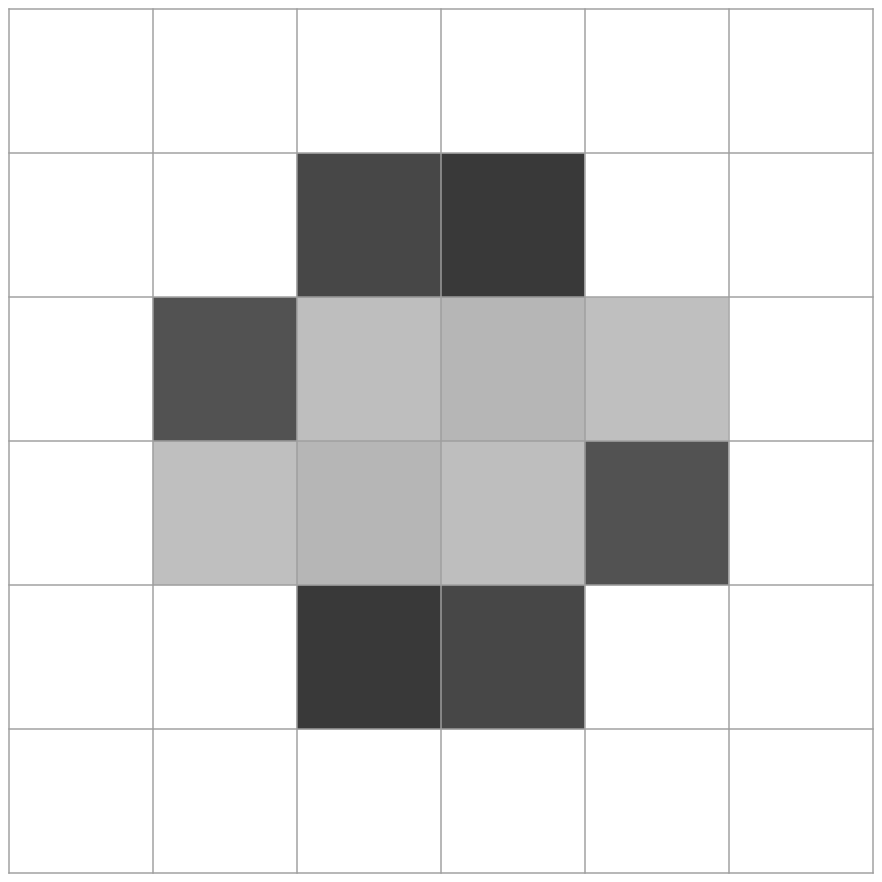}}}
\leftrightarrow
\vcenter{\hbox{\includegraphics[scale=0.25]{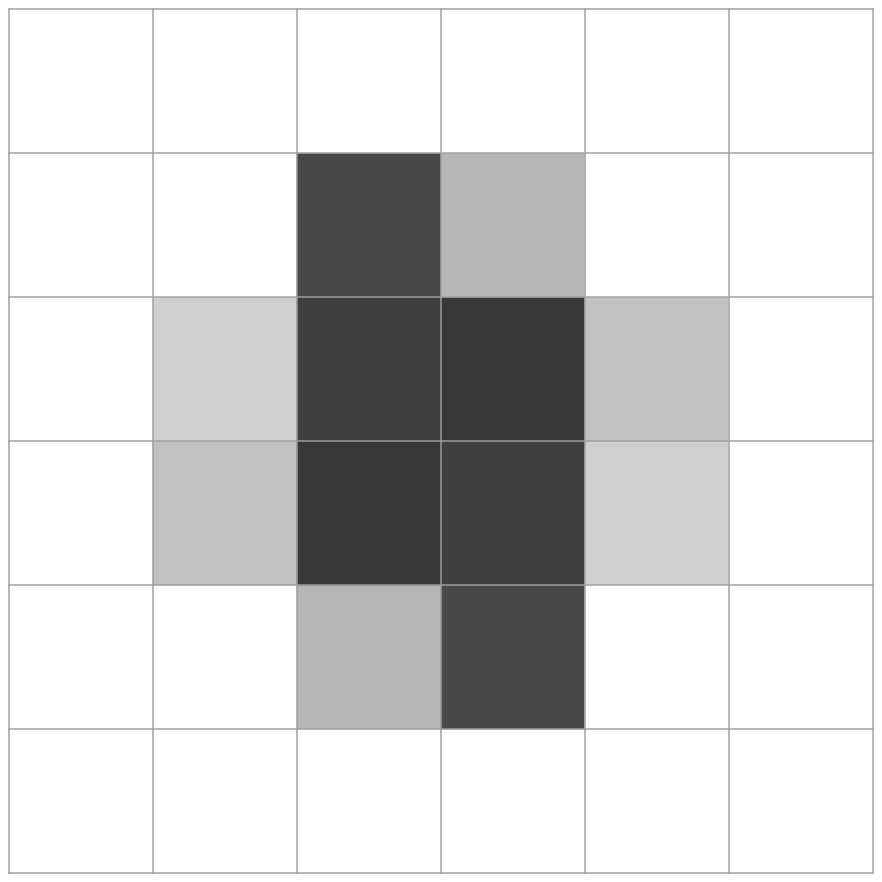}}}
\qquad
\medskip
\\
\text{(a) $a_{13}=0.25$} & \text{(b) $a_{13}=0.5$}
\medskip
\\
\vcenter{\hbox{\includegraphics[scale=0.25]{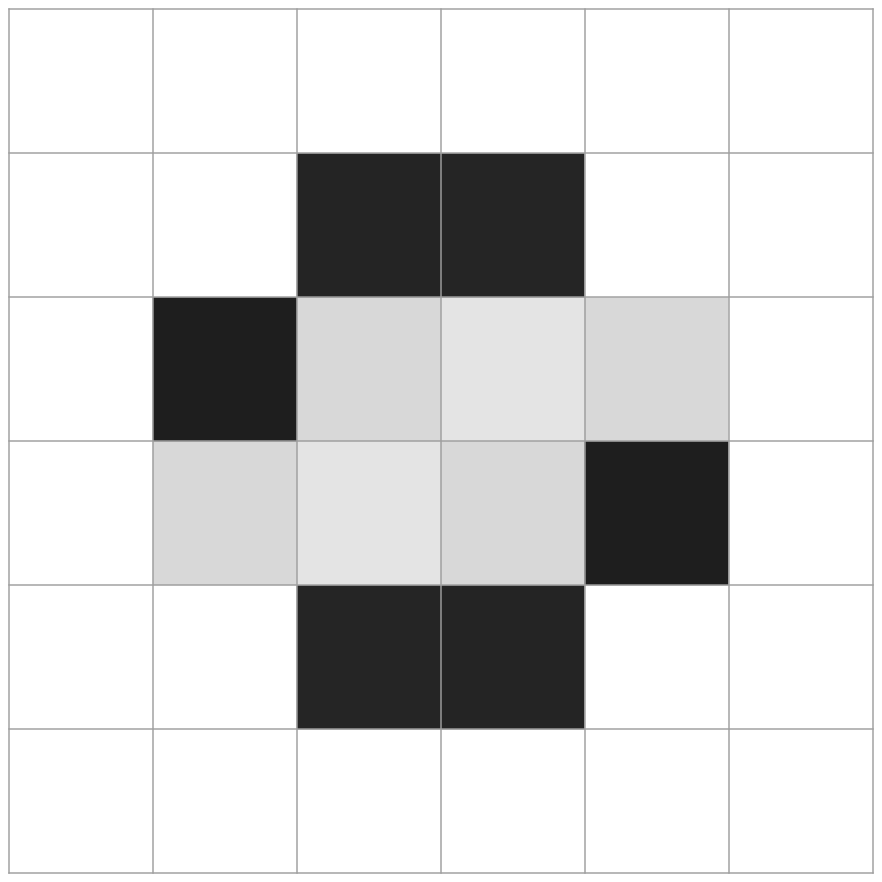}}}
\leftrightarrow
\vcenter{\hbox{\includegraphics[scale=0.25]{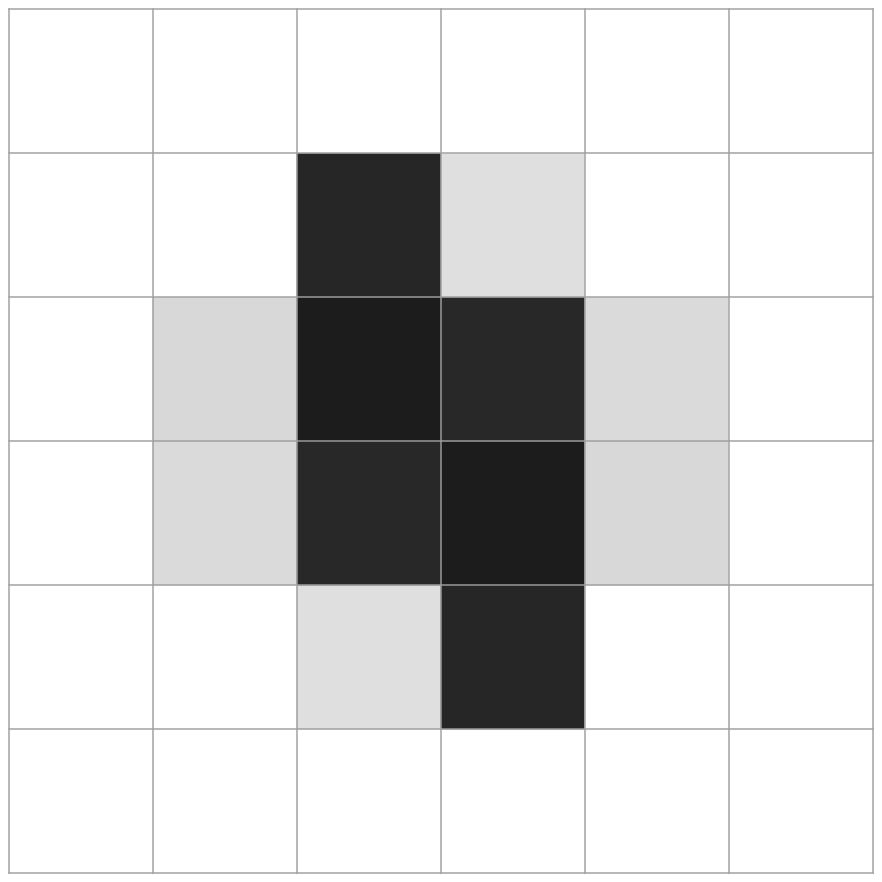}}}
&
\qquad
\vcenter{\hbox{\includegraphics[scale=0.25]{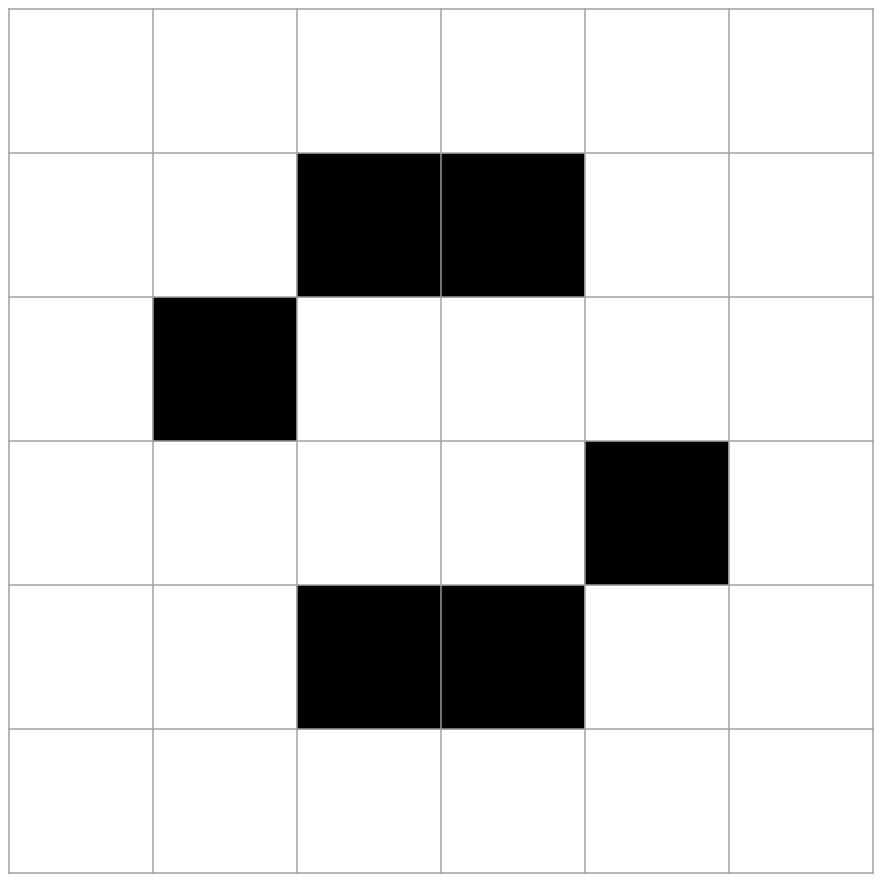}}}
\leftrightarrow
\vcenter{\hbox{\includegraphics[scale=0.25]{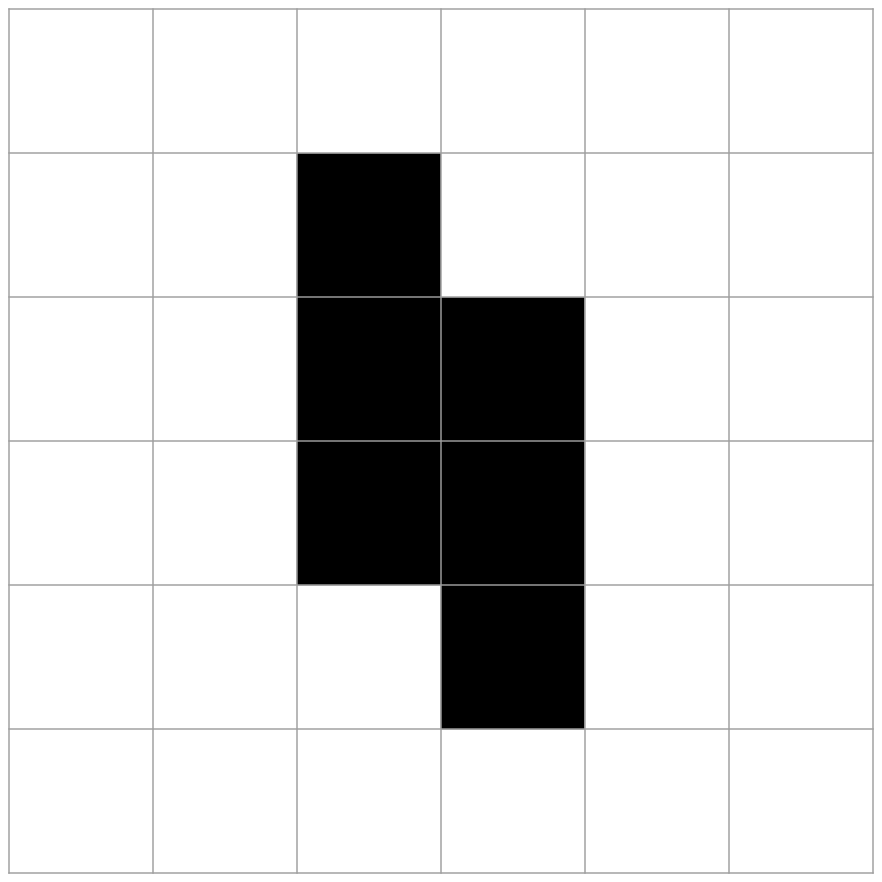}}}
\qquad
\medskip
\\
\text{(c) $a_{13}=0.75$} & \text{(d) $a_{13}=1$}
\end{array}$
\end{center}
\caption{Examples of solution by equation (\ref{u by a}).}
\label{fig:u by a}
\end{figure}

%%%%%%%%%%%%%%%%%%%%%%%%%%%%%%%%%%%%%%%%%%%%%%%%%%
There are other variations of solution obtained by rotating or reflecting those described in Subsections~\ref{subsec:clock} and \ref{subsec:toad}.  We can derive a general solution unifying all such solutions.  Assume a periodic and symmetric solution with period 2 and with a point symmetry confined in $4\times4$ region as shown below.
\begin{equation*}
\renewcommand{\arraycolsep}{2pt}
\begin{array}{cccccc}
0&0       &0       &0       &0       &0 \\
0&0       &u_{20}^0&u_{10}^0&0       &0 \\
0&u_{31}^0&u_{21}^0&u_{11}^0&u_{01}^0&0 \\
0&\boxed{u_{01}^0}&\boxed{u_{11}^0}&\boxed{u_{21}^0}&\boxed{u_{31}^0}&0 \\
0&0       &\boxed{u_{10}^0}&\boxed{u_{20}^0}&0       &0 \\
0&0       &0       &0       &0       &0 \\
\end{array}
\quad\leftrightarrow\quad
\begin{array}{cccccc}
0&0       &0       &0       &0       &0 \\
0&0       &u_{20}^1&u_{10}^1&0       &0 \\
0&u_{31}^1&u_{21}^1&u_{11}^1&u_{01}^1&0 \\
0&\boxed{u_{01}^1}&\boxed{u_{11}^1}&\boxed{u_{21}^1}&\boxed{u_{31}^1}&0 \\
0&0       &\boxed{u_{10}^1}&\boxed{u_{20}^1}&0       &0 \\
0&0       &0       &0       &0       &0 \\
\end{array}
\end{equation*}
Since periodicity and point-symmetry are assumed, boxed variables are only necessary to be determined.  Then the general solution is given by 20 parameters $a_i$ ($1\le i\le20$) as follows.
\begin{equation} \label{u by a}
\renewcommand{\arraycolsep}{0pt}
\begin{array}{rlllllllll}
u_{11}^0=a_{1}&+a_{3}&+a_{5}&+a_{8}&+a_{9}&+a_{12}&+a_{17}&+a_{18}&+a_{19}&+a_{20},\\
u_{21}^0=a_{1}&+a_{4}&+a_{6}&+a_{7}&+a_{10}&+a_{11}&+a_{17}&+a_{18}&+a_{19}&+a_{20},\\
u_{01}^0=a_{2}&+a_{4}&+a_{7}&+a_{8}&+a_{10}&+a_{12}&+a_{13}&+a_{15}&+a_{16}&+a_{20},\\
u_{10}^0=a_{2}&+a_{4}&+a_{5}&+a_{6}&+a_{9}&+a_{11}&+a_{13}&+a_{14}&+a_{16}&+a_{17},\\
u_{20}^0=a_{2}&+a_{3}&+a_{5}&+a_{6}&+a_{10}&+a_{12}&+a_{13}&+a_{14}&+a_{15}&+a_{18},\\
u_{31}^0=a_{2}&+a_{3}&+a_{7}&+a_{8}&+a_{9}&+a_{11}&+a_{14}&+a_{15}&+a_{16}&+a_{19},\\
u_{11}^1=a_{1}&+a_{3}&+a_{5}&+a_{8}&+a_{9}&+a_{12}&+a_{13}&+a_{14}&+a_{15}&+a_{16},\\
u_{21}^1=a_{1}&+a_{4}&+a_{6}&+a_{7}&+a_{10}&+a_{11}&+a_{13}&+a_{14}&+a_{15}&+a_{16},\\
u_{01}^1=a_{2}&+a_{4}&+a_{7}&+a_{8}&+a_{9}&+a_{11}&+a_{16}&+a_{17}&+a_{19}&+a_{20},\\
u_{10}^1=a_{2}&+a_{4}&+a_{5}&+a_{6}&+a_{10}&+a_{12}&+a_{13}&+a_{17}&+a_{18}&+a_{20},\\
u_{20}^1=a_{2}&+a_{3}&+a_{5}&+a_{6}&+a_{9}&+a_{11}&+a_{14}&+a_{17}&+a_{18}&+a_{19},\\
u_{31}^1=a_{2}&+a_{3}&+a_{7}&+a_{8}&+a_{10}&+a_{12}&+a_{15}&+a_{18}&+a_{19}&+a_{20},
\end{array}
\end{equation}
where $0\le a_i\le1$ for any $i$ and $a_1+a_2+\cdots+a_{20}=1$.  If we set $a_i=1$ and $a_j=0$ ($j\ne i$), one of the solutions reported in Subsections~\ref{subsec:clock} and \ref{subsec:toad} or its reflection or rotation is obtained.  Note that 20 parameters are redundant and we can reduce them to 6 through the transformation of parameters though the above expression is convenient to give a special solution.  Figure~\ref{fig:u by a} shows examples of solution obtained by equation (\ref{u by a}).  Figure~\ref{fig:u by a} (a), (b) and (c) show solutions for $a_{13}=0.25$, 0.5 and 0.75 respectively where other $a_i$'s are randomly given.  We can observe that the `toad' solution to GoL emerges as $a_{13}$ approaches 1.
%%%%%%%%%%%%%%%%%%%%%%%%%%%%%%%%%%%%%%%%%%%%%%%%%%
\section{MaxLife as multi-value cellular automaton}  \label{sec:multi-value}
\begin{figure}[hbtp]
\begin{center}
\begin{tabular}{ccc}
\includegraphics[scale=0.35]{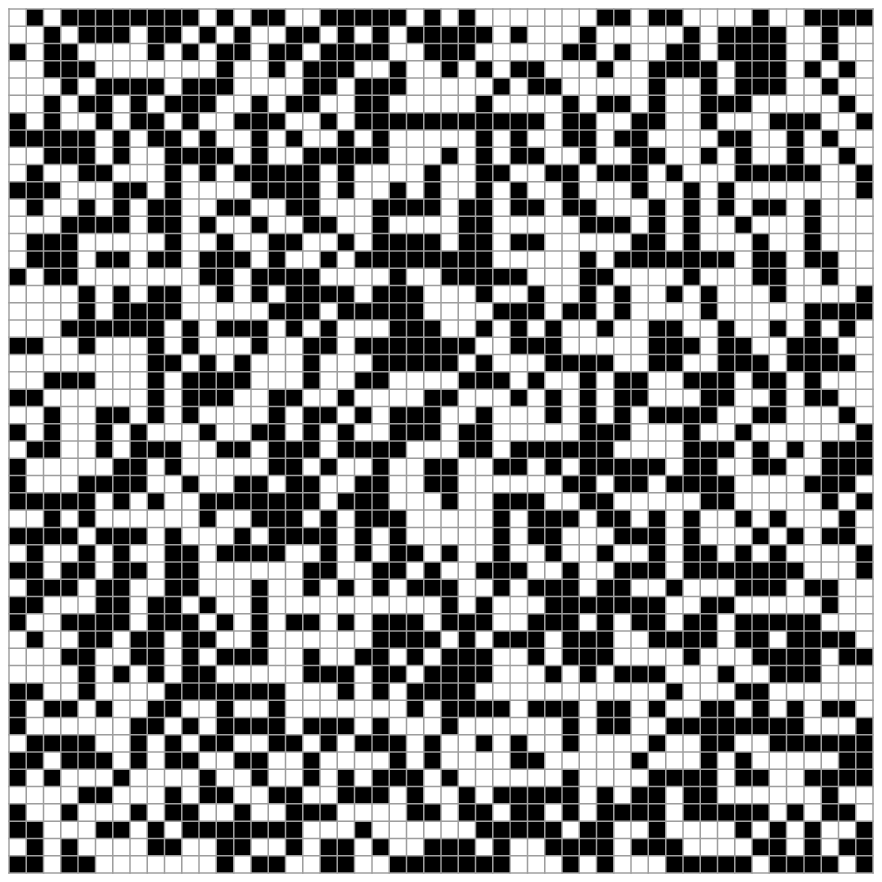}
&
\includegraphics[scale=0.35]{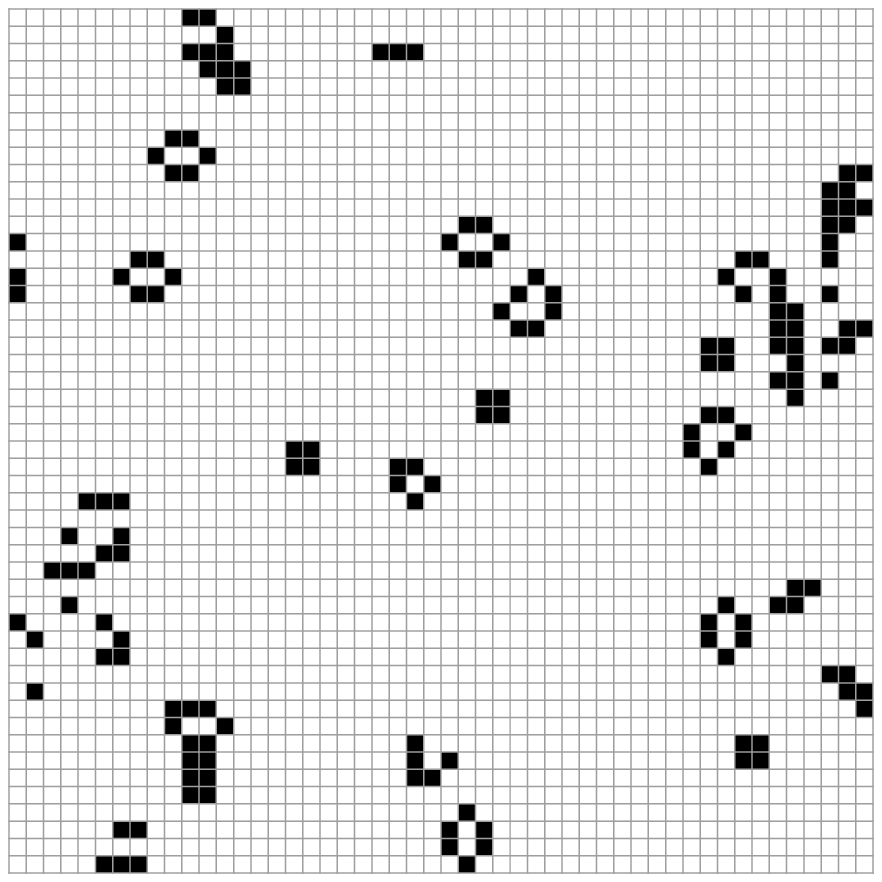}
&
\includegraphics[scale=0.35]{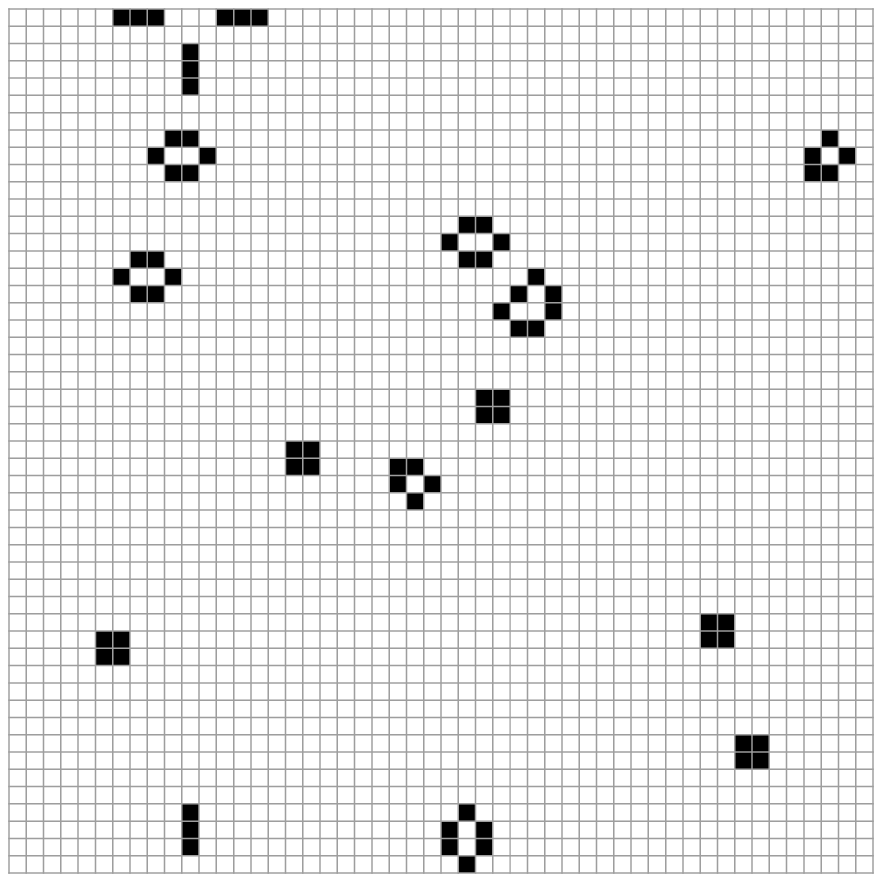}
\\
$n=0$ & $n=200$ & $n=400$ \\
\multicolumn{3}{c}{(a) 2-value case} \medskip\\
\includegraphics[scale=0.35]{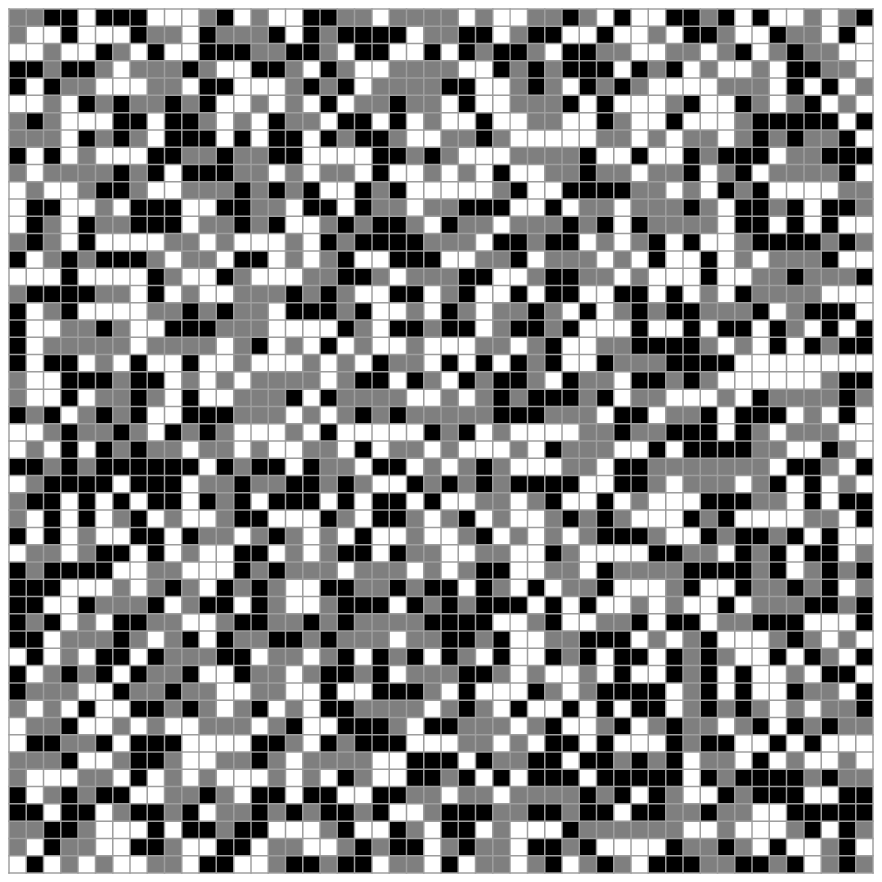}
&
\includegraphics[scale=0.35]{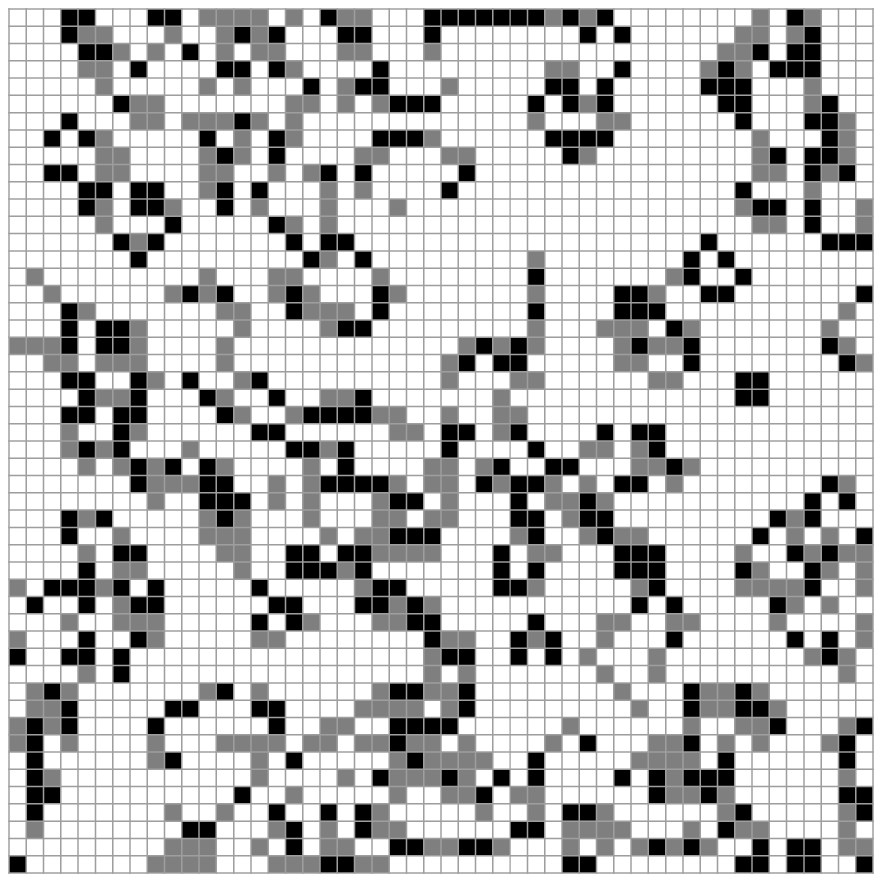}
&
\includegraphics[scale=0.35]{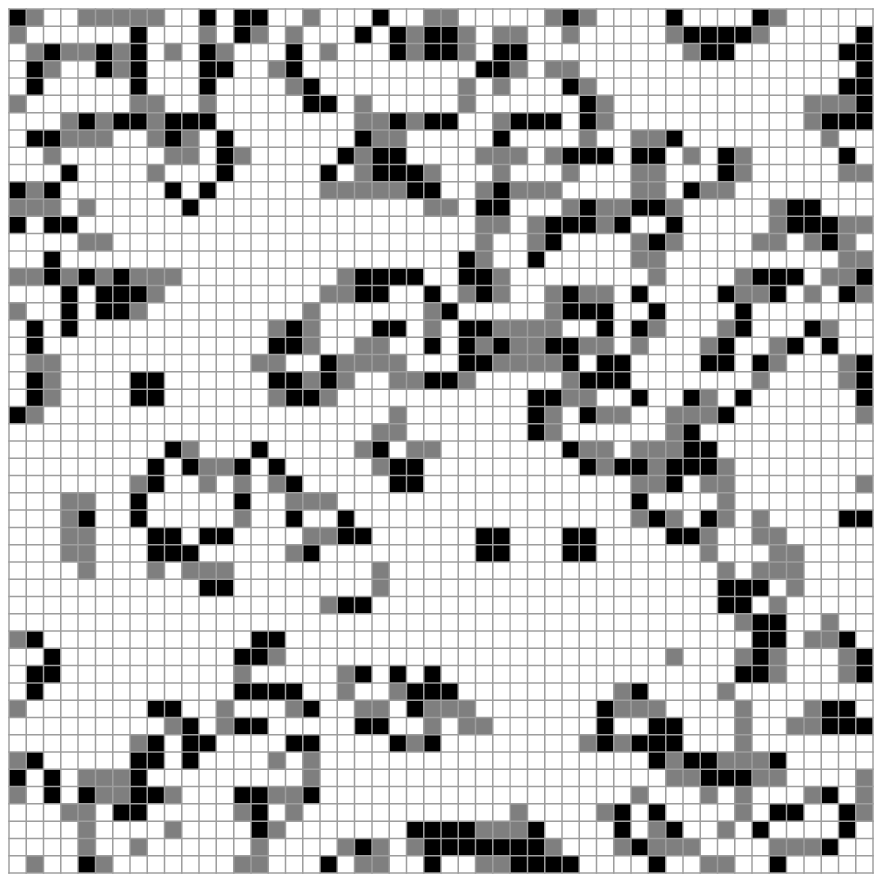}
\\
$n=0$ & $n=200$ & $n=400$ \\
\multicolumn{3}{c}{(b) 3-value case} \medskip\\
\includegraphics[scale=0.35]{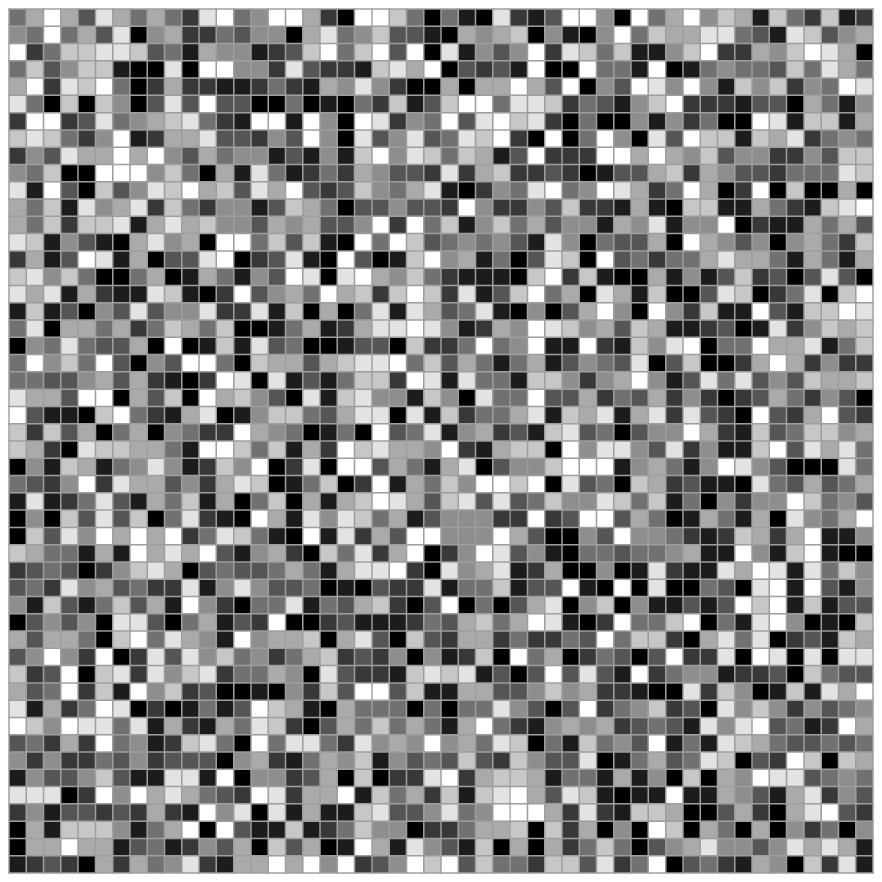}
&
\includegraphics[scale=0.35]{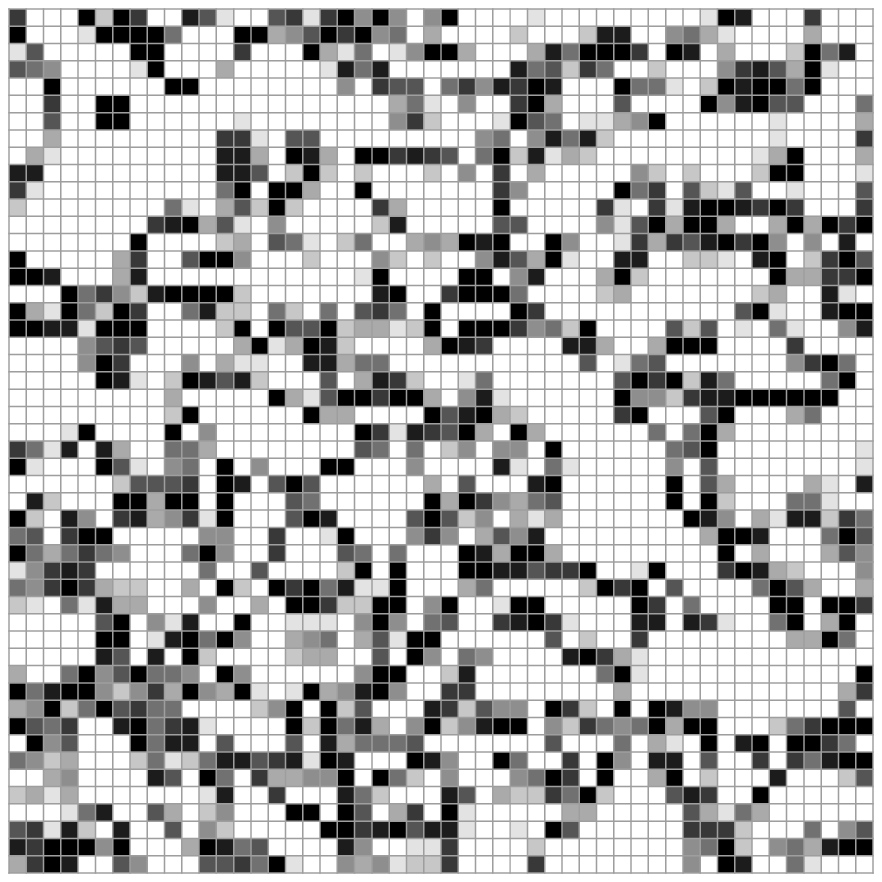}
&
\includegraphics[scale=0.35]{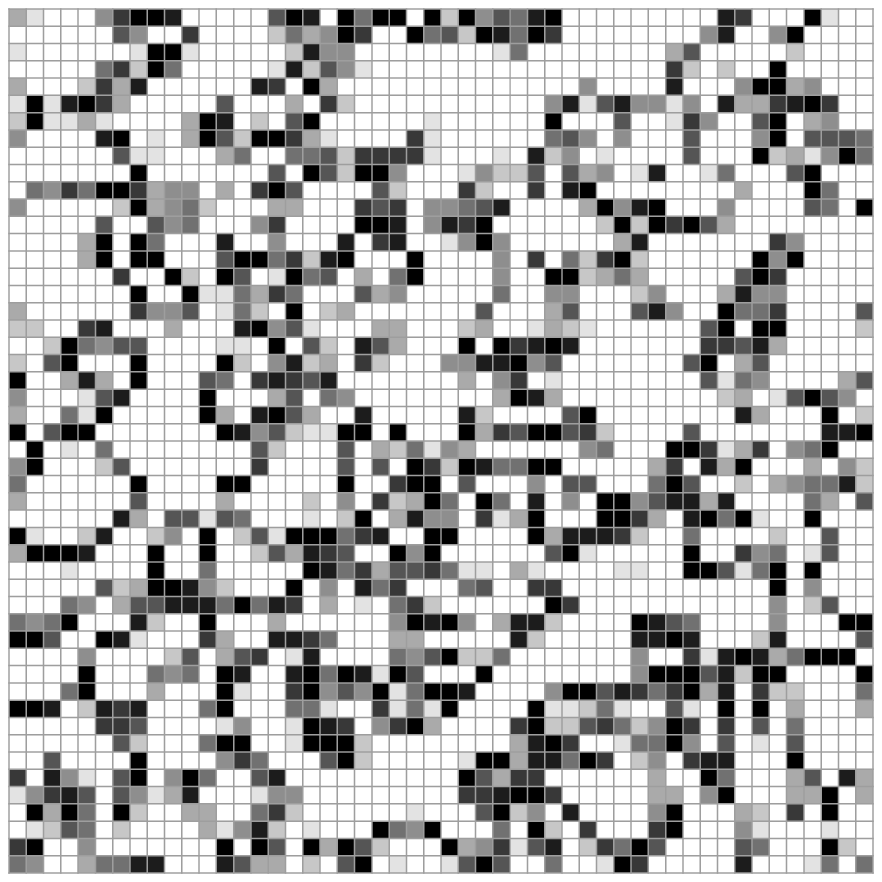}
\\
$n=0$ & $n=200$ & $n=400$ \\
\multicolumn{3}{c}{(c) 10-value case}
\end{tabular}
\end{center}
\caption{Evolution from random initial data for equation (\ref{maxlife}) as multi-value CA.}
\label{fig:multi-value}
\end{figure}

%%%%%%%%%%%%%%%%%%%%%%%%%%%%%%%%%%%%%%%%%%%%%%%%%%
The range of $u$ in equation (\ref{maxlife}) can be closed to the finite set $\{0, 1/N,$\break
$2/N, \ldots, (N-1)/N,1\}$ for a positive integer $N$.  Then equation (\ref{maxlife}) becomes ($N+1$)-value CA.  The case $N=1$ is the original GoL.  Figure~\ref{fig:multi-value} shows an example of evolution for 2, 3 and 10-value cases ($N=1$, 2, 9) with periodic boundary condition from random initial data.
Since it is difficult to evaluate the behavior of general solution quantitatively, we describe our observation from numerical computation as shown in Figure~\ref{fig:multi-value}.  Solution of 2-value case (GoL) tends to change drastically as time proceeds and often results in the steady state with separated static and periodic patterns.  In contrast to 2-value case, solution of 3 or more value case rarely results in the steady state and continues to evolve with connected non-zero domains interacting one another.\par
  There are various basic static or periodic solutions confined in a finite region for multi-value case.  Some of them can be obtained by choosing the parameters of solutions reported in Section~\ref{sec:solutions}.  For example, if $a$ is set to $1/2$ for the solutions shown in Figures~\ref{fig:max blinker}, \ref{fig:max clock}, \ref{fig:max toad} and \ref{fig:max glider}, they become solutions for 3-value case, and if $1/3$, 4-value case.  Moreover, if the dimensions of region for $u\ne0$ and the period are assumed, all solutions can be searched numerically.  For example, there are 40 static solutions and 23 periodic solutions with period 2 for 3-value case.  Solutions to GoL are included in them and 13 of 40 static solutions and 3 of 23 periodic solutions are constructed only from 0 and 1.  The number of steady basic solutions for 3-value case are much larger than 2-value case, and it suggests the persistence of evolution of non-zero area for multi-value case as shown in Figure~\ref{fig:multi-value}.
%%%%%%%%%%%%%%%%%%%%%%%%%%%%%%%%%%%%%%%%%%%%%%%%%%
\section{Concluding remarks}  \label{sec:conclusion}
%%%%%%%%%%%%%%%%%%%%%%%%%%%%%%%%%%%%%%%%%%%%%%%%%%
  We proposed the max-plus equation (\ref{maxlife}) as the difference equation on a real-valued state variable.  It includes GoL as a special case if the state value is restricted to 0 and 1.  It has special solutions including a free parameter and they unify those to GoL by special choice of parameter.  Though various solutions to GoL have been reported independently, their relations are suggested through this unification.  Among such solutions, we obtained a solution (\ref{u by a}) including many parameters unifying various solutions to GoL.  However, a systematic way to derive general solutions has not been found yet. It is one of future problems to propose a way to solve equation (\ref{maxlife}) as we solve the differential equations.\par
  Max-plus equation can be obtained from difference equation using an exponential type of transformation of variables with a limiting parameter.  Consider the following difference equation,
\begin{equation}  \label{dlife}
  U_{ij}^{n+1}=C\frac{(1+\delta^2U_{ij}^nS_{ij}^n)(1+\delta^4S_{ij}^n)}{(1+\delta^3U_{ij}^nS_{ij}^n)(1+\delta^3S_{ij}^n)},
\end{equation}
where
\begin{equation*}
  S_{ij}^n=
  U_{i-1j-1}^nU_{ij-1}^nU_{i+1j-1}^nU_{i-1j}^nU_{i+1j}^nU_{i-1j+1}^nU_{ij+1}^nU_{i+1j+1}^n,
\end{equation*}
and
\begin{equation*}
  C=\frac{(1+\delta^3)^2}{(1+\delta^2)(1+\delta^4)}.
\end{equation*}
If we use the transformation including a parameter $\varepsilon$,
\begin{equation*}
  U_{ij}^n=e^{u_{ij}^n/\varepsilon},\qquad S_{ij}^n=e^{s_{ij}^n/\varepsilon},\qquad \delta=e^{-1/\varepsilon},
\end{equation*}
equation (\ref{maxlife}) is obtained from equation (\ref{dlife}) by the limit $\varepsilon\to+0$.  Note that the limit formula (\ref{ultra}) is used in the derivation.\par
\begin{figure}[hbtp]
\begin{center}
\begin{tabular}{ccc}
\includegraphics[scale=0.35]{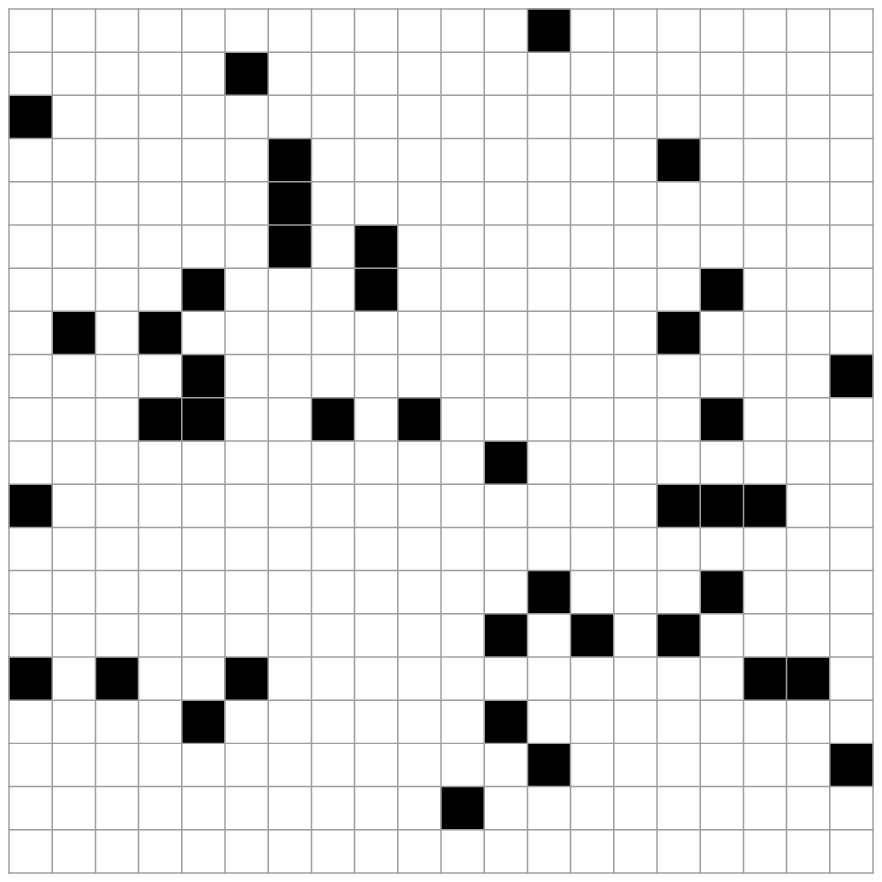}
&
\includegraphics[scale=0.35]{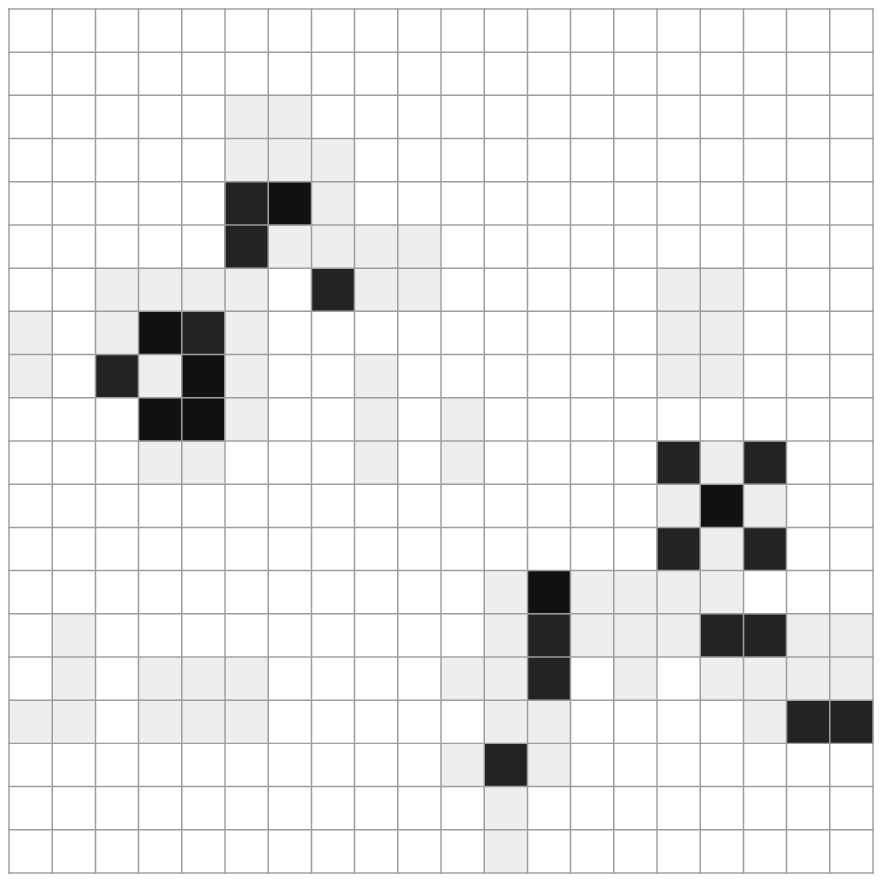}
&
\includegraphics[scale=0.35]{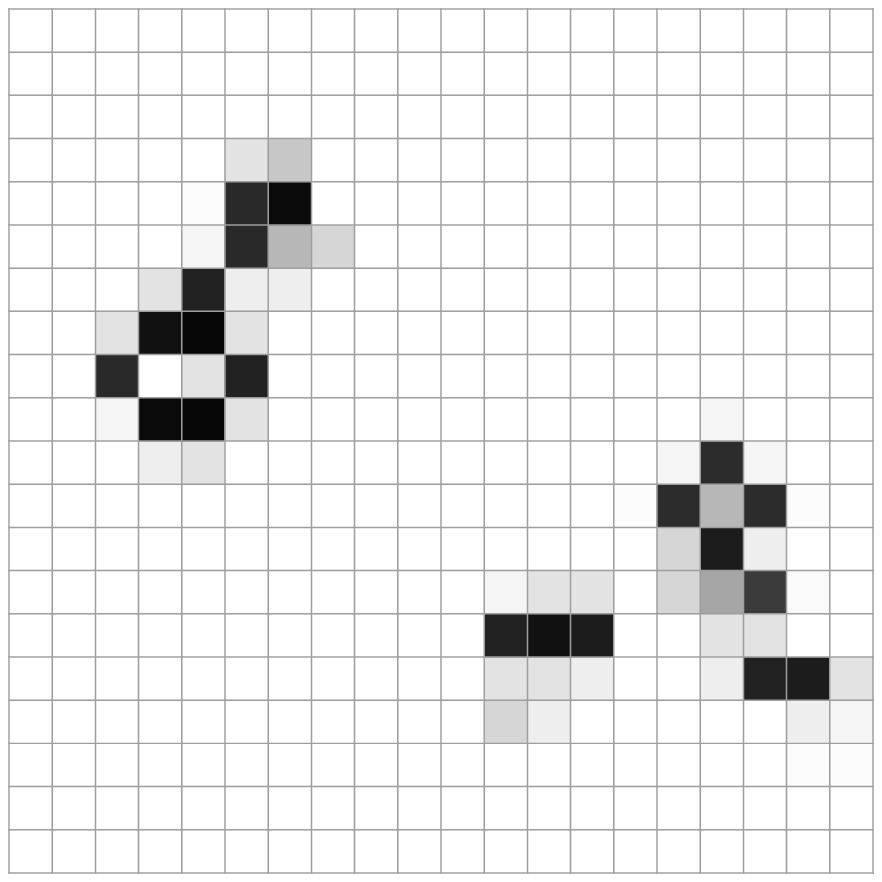}
\\
$n=0$ & 1 & 2 \medskip\\
\includegraphics[scale=0.35]{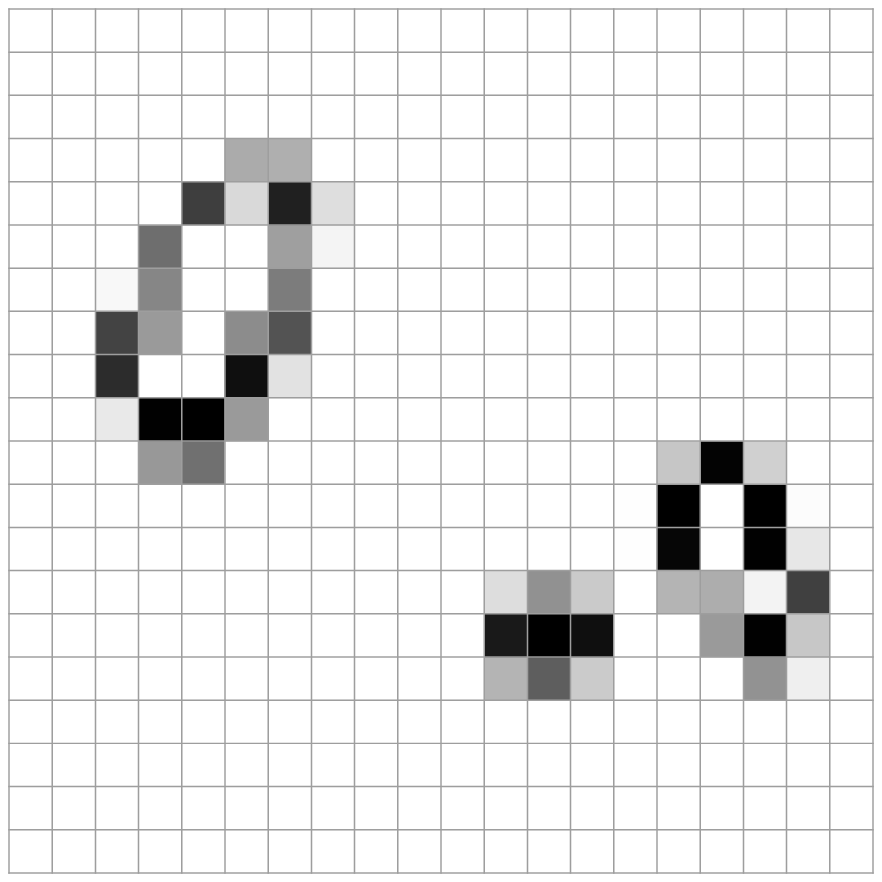}
&
\includegraphics[scale=0.35]{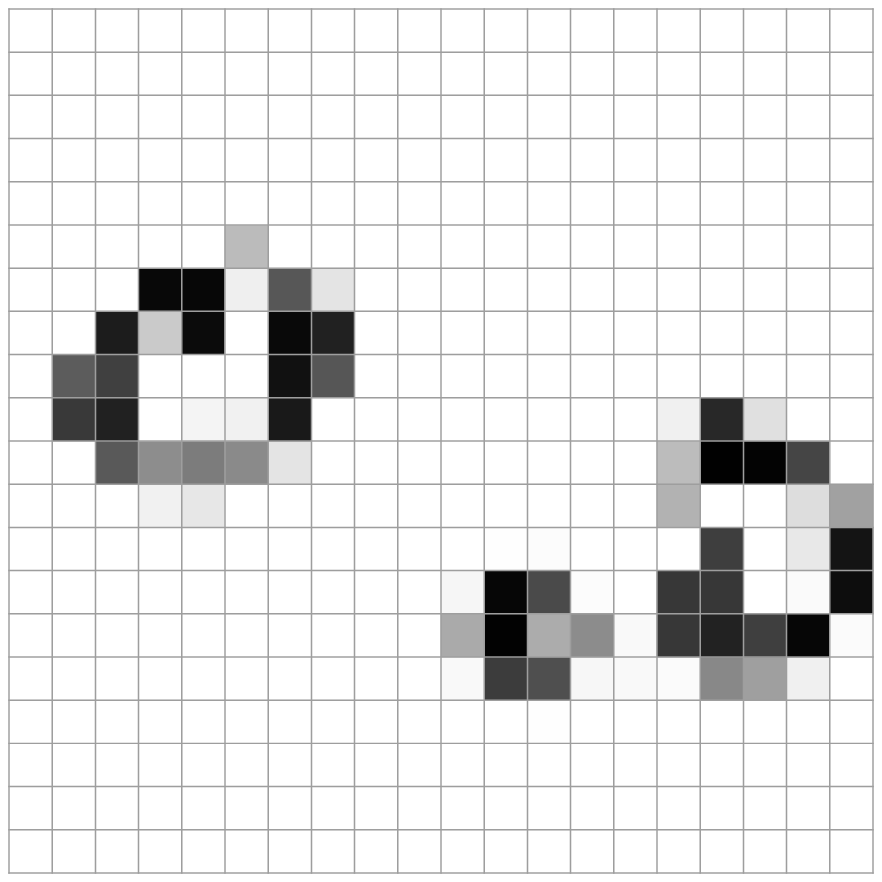}
&
\includegraphics[scale=0.35]{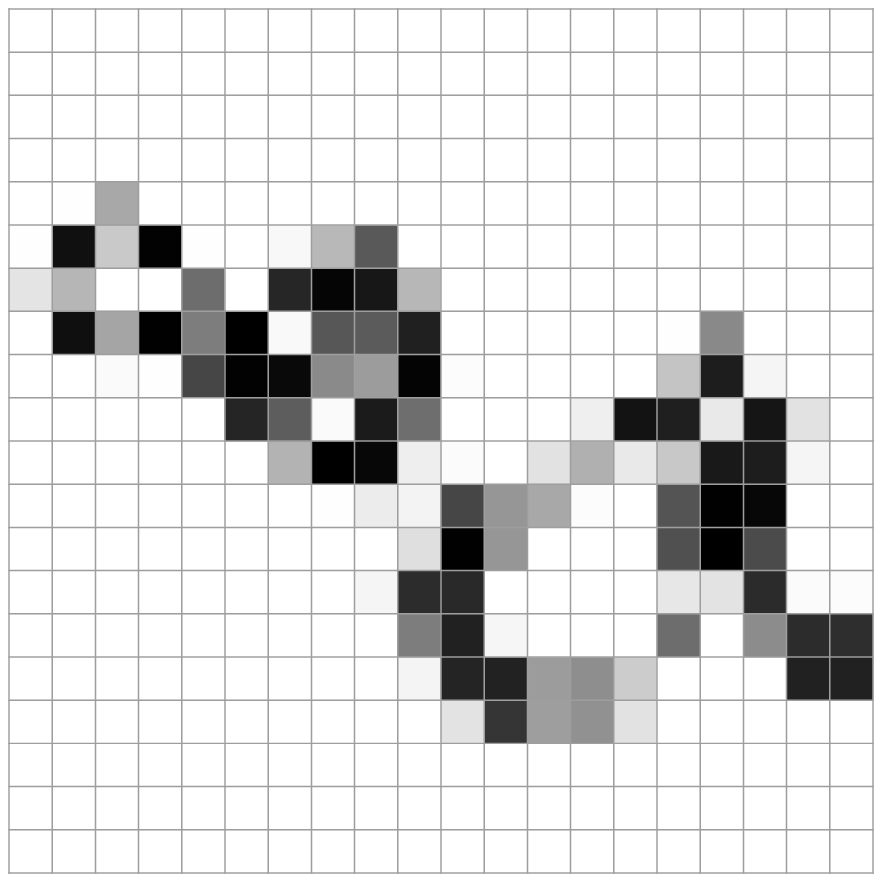}
\\
4 & 8 & 16 \medskip\\
\includegraphics[scale=0.35]{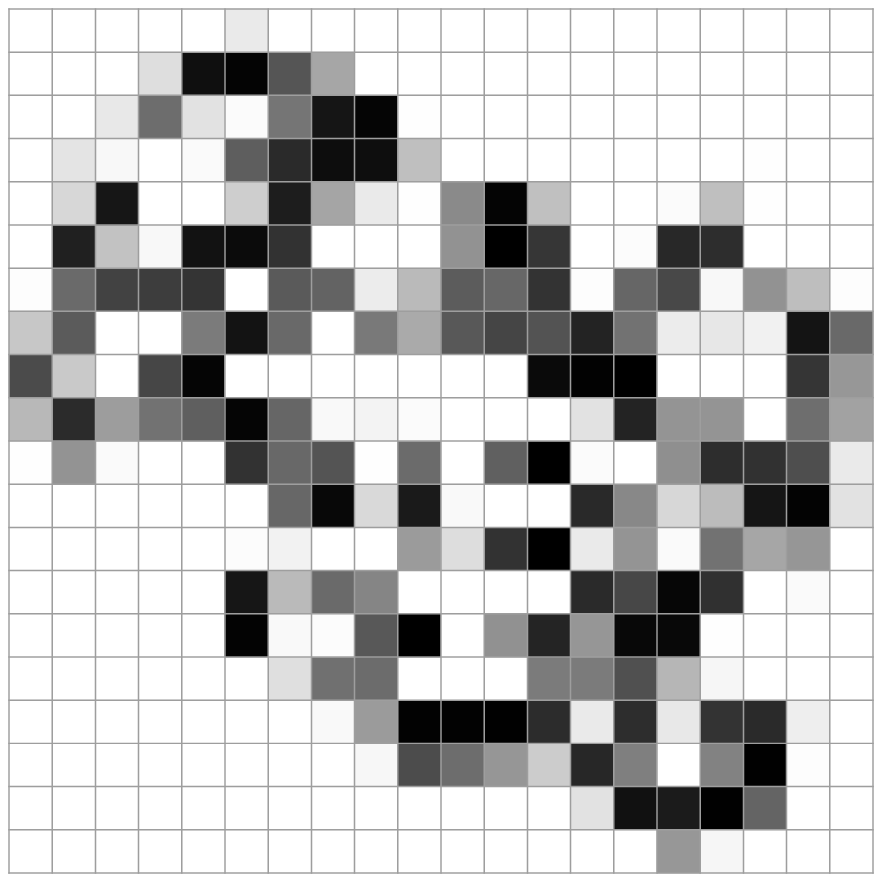}
&
\includegraphics[scale=0.35]{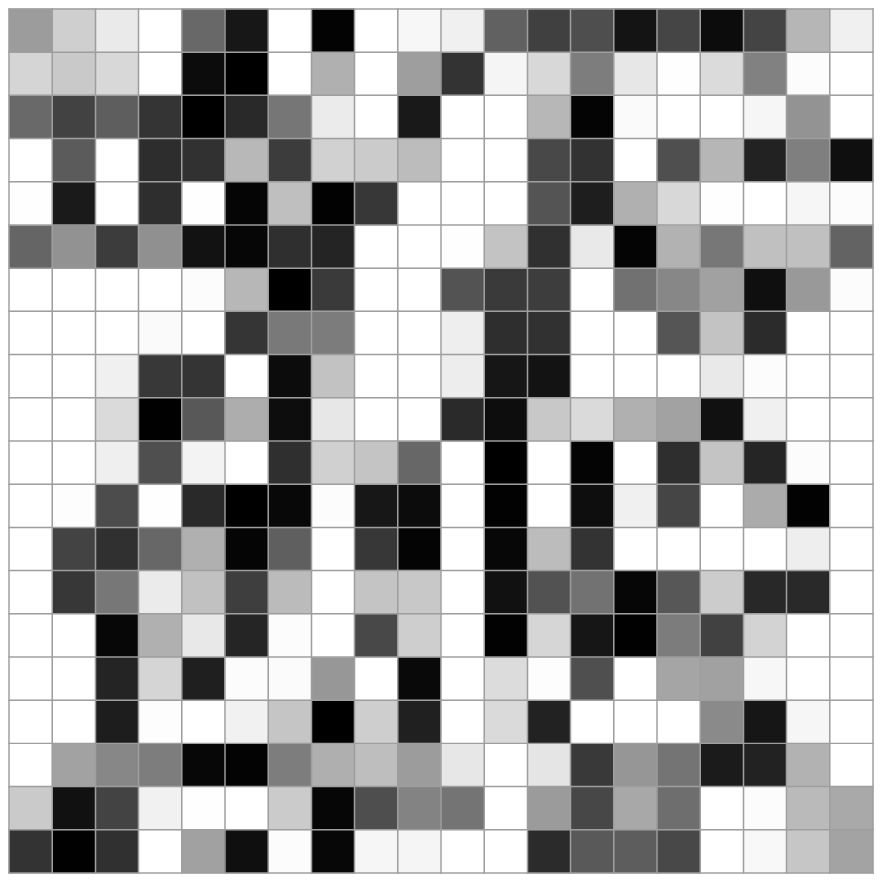}
&
\includegraphics[scale=0.35]{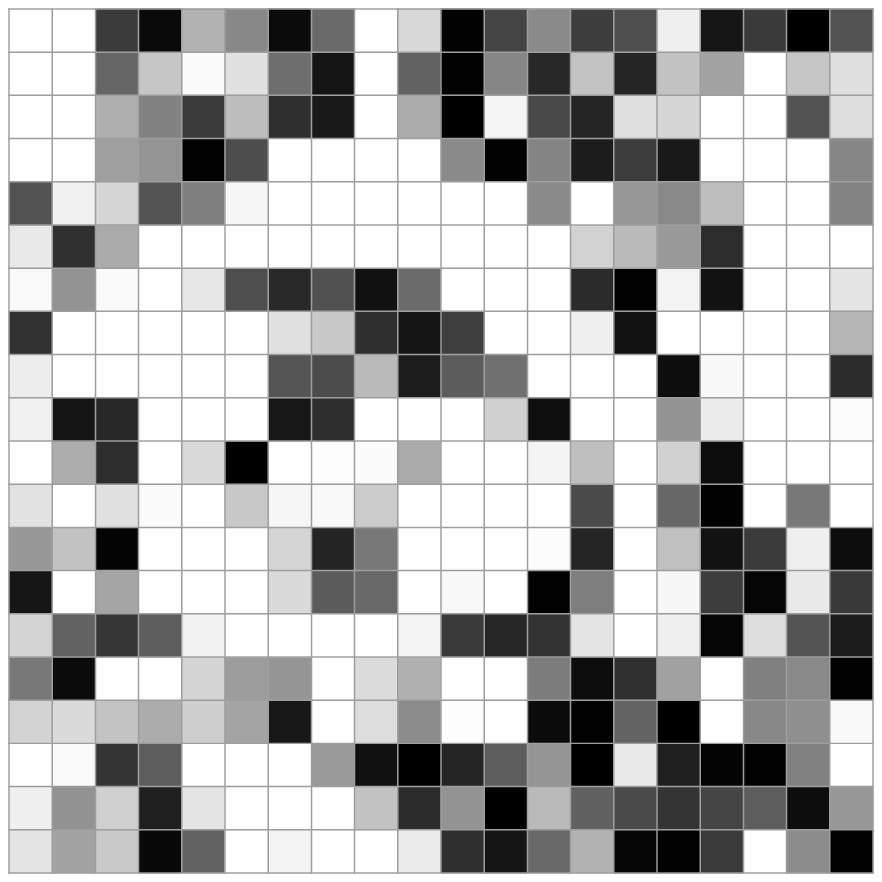}
\\
32 & 64 & 128
\end{tabular}
\end{center}
\caption{Example of evolution of solution to equation (\ref{dlife}) for $\varepsilon=0.1$.}
\label{fig:dlife}
\end{figure}

  Example of evolution of solution to equation (\ref{dlife}) is shown in Figure~\ref{fig:dlife} for $\varepsilon=0.1$.  The background value is $U=e^{0/\varepsilon}=1$ and randomly chosen cells are set to $U=e^{1/\varepsilon}$ for initial data.  The initial data changes drastically at $n=1$, some patterns survive and evolve from $n=2$ to 16, and they merge and extend to the whole area from $n=32$ to 128.  This extended pattern is evolving until at least $n=10000$ and the range of $U$ is always preserved about from 1 to $e^{1/\varepsilon}$.  Though some stable static patterns confined in a finite area are found numerically, exact solutions giving static, periodic or moving patterns have not yet been found.  It is another future problem to find exact solutions and to discuss the relation between solutions to equation (\ref{maxlife}) and to equation (\ref{dlife}).
%%%%%%%%%%%%%%%%%%%%%%%%%%%%%%%%%%%%%%%%%%%%%%%%%%

\end{document}